\documentclass[12pt,aps,prd,preprint,tightenlines,superscriptaddress,
  showpacs,nofootinbib,floatfix]{revtex4}
\newcommand{\PRE}[1]{{#1}} 

\usepackage{hyperref,rotating}      
\usepackage{bm}
\usepackage{amsmath}
\usepackage{graphicx}
\usepackage{array}
\usepackage{subfigure}
\renewcommand{\eqref}[1]{Eq.~(\ref{#1})}

\newcommand{\Eqref}[1]{Equation~(\ref{#1})}

\newcommand{\bea}{\begin{eqnarray}}
\newcommand{\eea}{\end{eqnarray}}
\newcommand{\beq}{\begin{equation}}
\newcommand{\eeq}{\end{equation}}
\newcommand{\beqa}{\begin{eqnarray}}
\newcommand{\eeqa}{\end{eqnarray}}
\newcommand{\nn}{\nonumber}

\def\GeV{\mathrm{GeV}}
\def\TeV{\mathrm{TeV}}
\begin{document}



\title{ \PRE{\vspace*{1.5in}} Flavored Gauge Mediation,
A Heavy Higgs, and Supersymmetric Alignment
\\
 }

\author{Mohammad Abdullah\PRE{\vspace*{.1in}}}
\affiliation{Department of Physics and Astronomy, University of
California, Irvine, CA 92697, USA
\PRE{\vspace*{.1in}}}

\author{Iftah Galon\PRE{\vspace*{.1in}}}
\affiliation{Physics Department, Technion---Israel Institute of
Technology,\\ Haifa 32000, Israel
\PRE{\vspace*{.3in}}}

\author{Yael Shadmi\PRE{\vspace*{.1in}}}
\affiliation{Physics Department, Technion---Israel Institute of
Technology,\\ Haifa 32000, Israel
\PRE{\vspace*{.3in}}}
\author{Yuri Shirman}
\affiliation{Department of Physics and Astronomy, University of
California, Irvine, CA 92697, USA
\PRE{\vspace*{.1in}}}


\begin{abstract}
\PRE{\vspace*{.3in}} 
We show that the messenger-matter couplings of Flavored Gauge
Mediation Models can generate substantial stop mixing
and new contributions to the stop masses,
 leading
to Higgs masses around 126~GeV with 
sub-TeV superpartners, and with some colored superpartners
around 1-2~TeV in parts of the parameter space.
We study the spectra of a few examples with a single messenger
pair coupling dominantly  to the top, for different messenger scales.
Flavor constraints in these models are obeyed by virtue of 
supersymmetric alignment: the same flavor symmetry that explains
fermion masses dictates the structure of the matter-messenger couplings,
and this structure is inherited by the soft terms.
We  
present the leading 1-loop and 2-loop contributions to the soft
terms for general coupling matrices in generation space. 
Because of the 
Higgs-messenger mixing induced by the new couplings,
the calculation of these soft terms via
analytic continuation requires careful matching of the
high- and low-energy theories. 
We discuss the calculation in detail in the Appendix. 
\end{abstract}


\maketitle

\section{Introduction}
\label{sec:introduction}
The null results of direct searches for supersymmetry
suggest that it does not manifest itself in the form of light, 
flavor-blind superpartners.
Furthermore, if the recently discovered  scalar at 126~GeV~\cite{atlas,cms}
is indeed the Higgs boson, its relatively large mass
requires 
either a large stop mixing or very heavy stops,
at least in the context of the Minimal Supersymmetric Standard Model
(MSSM)~\cite{Carena:2002es}.
These results are especially problematic for 
Gauge-Mediated Supersymmetry Breaking (GMSB) 
models~\cite{earlygmsb,dnns}, 
which predict zero $A$-terms 
and flavor-blind spectra at the messenger scale. 
Low scale gauge mediation is therefore strongly disfavored by
the Higgs mass, and
even high-scale models, with  $A$-terms generated 
by the running below the messenger scale,
require stop masses of around 8--10~TeV~\cite{Ajaib:2012vc,Feng:2012rn}
in the context of minimal gauge mediation~\cite{dnns}.
Given the tight relations between squark and gluino masses in minimal GMSB
(mGMSB), this implies that all 
superpartners are very heavy in these
models, and beyond the reach of any foreseeable experiment.

From a purely theoretical point of view, however, 
GMSB models are very attractive, 
since both the breaking of supersymmetry and its mediation are 
described by well understood quantum field theories,
as opposed to unknown Planck-scale physics.
Indeed, flavor-blind extensions of gauge mediation 
have been extensively discussed in recent years~\cite{Meade:2008wd}.
These extensions too are only consistent with a 126~GeV Higgs
for high messenger scales, unless the stops or the gluino
are very heavy~\cite{Draper:2011aa}.
Here we will study instead a flavor-dependent extension of gauge mediation,
specifically, the  Flavored Gauge Mediation 
(FGM) models of~\cite{Shadmi:2011hs}.
In these models,  the flavor structure of GMSB is in principle modified,
due to superpotential couplings of the messengers to SM 
fields~\cite{Dine:1996xk}. 
We will show that these messenger-matter couplings can 
yield  significant stop $A$-terms\footnote{Scenarios where non-trivial flavor structure leads to maximal flavor mixing have also been considered in the context of horizontal symmetries and MSSM, see for example \cite{Badziak:2012rf}}, as well as new contributions to
the stop soft masses, resulting  in  a heavy Higgs 
and fairly light superpartners for a wide range of  messenger scales.

The superpartner masses in FGM models are generated by both the SM 
gauge interactions, and by Yukawa-type superpotential couplings of 
the messengers to the SM matter fields.
Thus, while the interactions mediating the breaking are not purely gauge
interactions, they are still completely ``visible''--- occurring within
simple field theoretic extensions of the 
MSSM, and potentially at low scales. 
Since the matter-messenger couplings are in principle flavor dependent, 
they are strongly constrained by the non-observation of 
flavor changing neutral currents. 
As stressed in~\cite{Shadmi:2011hs}, however, there are good reasons
to consider these couplings. 
At the very least, given our ignorance about the origin 
of the SM Yukawas structure, it is conceivable that the 
messenger-matter couplings have some special structure, which results
in an acceptable pattern of soft terms. Indeed, the structure of the
SM fermion masses hints at some theory of flavor, and any such theory
will necessarily also control the sizes of matter-messenger couplings. 
Furthermore, as superpartner masses are pushed to higher values
by direct searches (as well as by the large Higgs mass), there is more
room for non-degenerate spectra.
>From the point of view of LHC searches
for supersymmetry, the assumption of 
Minimal Flavor Violation (MFV),
which  underlies many analyses, can result in 
reduced sensitivity to non-MFV 
spectra~\cite{Feng:2007ke,Galon:2011wh,Shadmi:2012sy}.
So when searching for gauge-mediated supersymmetry, it is important to keep 
the possibility of flavor-dependent spectra in mind,  
and FGM models  provide useful examples of such spectra.

The main new ingredient in our models will be superpotential 
messenger-matter interactions, 
with the up-type Higgs (or also the down-type
Higgs) replaced by a messenger field of the same gauge charges.
Since we would like to generate a large top $A$-term,
the messengers need only couple to the top.
As a concrete realization of this scenario, we invoke a flavor symmetry
under which the Higgs and messenger field have identical charges.
Flavor constraints in our models are thus satisfied by a combination of
degeneracy --- coming from the pure gauge contributions, 
and alignment~\cite{Shadmi:2011hs}. 
Unlike in the original alignment models of~\cite{Nir:1993mx},
in which the flavor symmetry controls the soft terms directly,
here it controls the {\sl supersymmetric} messenger-matter couplings
so that they are aligned with the SM Yukawas.
The soft terms therefore inherit this structure, even though
they are generated at the messenger scale, which is typically
much lower than the flavor-symmetry breaking scale.

We note that three other papers appeared 
recently~\cite{Albaid:2012qk,Kang:2012ra,Craig:2012xp}
which rely on messenger-SM couplings to raise the Higgs mass. 
The differences between our models and the models 
of~\cite{Albaid:2012qk,Kang:2012ra,Craig:2012xp} arise due to 
the different choices of symmetries and, as a result, the allowed 
messenger-SM couplings. 
In~\cite{Albaid:2012qk}, the messengers are chosen to have the same R-parity
as the SM matter fields, so the relevant coupling is the analog of the
Yukawa coupling with one SM matter field replaced by a messenger.
Messengers in 5$+\bar5$ representations of $SU(5)$ in these models 
do not affect the $u^c$ mass,
and as a result can raise the Higgs mass only to 
around 118~GeV for stops below 1.5~TeV. 
Therefore~\cite{Albaid:2012qk}
uses messengers in 10$+\overline{10}$.
In~\cite{Kang:2012ra,Craig:2012xp},
a coupling of the type Higgs-messenger-messenger
is used, with one messenger being a SM gauge singlet.
Since none of the fields involved is colored, the effect
of this coupling is moderate, so that the Higgs mass is viable
only at low messenger scales, where the negative one-loop contributions
to the Higgs soft  masses-squared are important~\cite{Craig:2012xp}.
It is interesting that, although
the new messenger couplings in our models do not involve the Higgs
at all, they have a significant effect on the Higgs mass.
The reason is that the key feature needed for getting a large Higgs
mass is a modified stop spectrum, which
only requires that the messengers couple to the top.

To calculate the 2-loop contributions to the soft terms we use analytic 
continuation~\cite{Giudice:1997ni}. The messenger-Higgs mixing present in the models
has led to some erroneous results in the literature (including in an earlier
version of this 
article\footnote{Specifically, our results for the Higgs masses were correct,
but our results for the remaining scalars were wrong. 
We thank M.~Ibe for pointing this out to us.}). 
We clarify this issue and discuss the calculation
in detail in the Appendix. 
The key  point is the correct identification
of the relevant couplings and wave function renormalizations,
for which one must match the high-energy and low-energy theories correctly.
We give a simple and intuitive prescription for the calculation by
identifying the physical messenger field, given by the heavy combination,
and the physical Higgs field, given by the light combination,
and their respective running couplings.

The resulting spectra are rich and quite unusual.
There are essentially two different regions of the parameter space
which lead to an enhanced Higgs mass.
In the first region, the stop A-terms are substantial, while
the LL and RR stop masses are largely unmodified, because they receive
both positive and negative contributions from the new couplings.
Since these negative contributions are dominated by 1-loop effects
which are higher order in the supersymmetry breaking, this region
occurs for relatively low messenger scales.
The resulting spectra can be very light, with colored superpartners even 
well below 2~TeV.
In the second region, the LL and RR stop masses are enhanced as well,
so that the large Higgs mass is driven not by the large mixing but rather
by large stop masses. Thus a 125~GeV Higgs requires heavy stops.
While only the stop masses are modified at the messenger scale,
the running to the weak scale can affect the remaining masses dramatically,
since the stop masses are  large. While most colored superpartners
are above 2~TeV in this case, the weak gauginos and sleptons
can be light. Furthermore, the new contributions to the soft masses
can reverse the effect of the stops on the remaining sfermions
through the RGEs, leading to novel spectra with the NLSP being either
the neutralino, or a L-handed slepton.

The organization of this paper is as follows. 
In Section~\ref{sec:FGM} we briefly review FGM and introduce the 
symmetries and the superpotential of our models. 
In Section \ref{sec:soft terms}  we give expressions for 
the soft terms in the limit of third-generation 
dominance.
In Section \ref{sec:phenom} we present the Higgs mass and superpartner 
spectra for different choices of parameters. 
Our conclusions and discussion of the results are presented in 
Section~\ref{sec:conclusions}.
We present the calculation of the soft masses in the Appendix.
In~\ref{analcontapp}, we derive the soft terms in a simple toy model with Higgs-messenger 
mixing. We generalize this to the models of interest in~\ref{multiple}.
Full expressions for the soft terms for general 
$3\times 3$ coupling matrices in generation space are presented in~\ref{final}. 
As a cross-check, we also calculated the relevant terms explicitly.
We describe this calculation in~\ref{explicitcal}.

\section{Models}
\label{sec:models}
\subsection{The models and supersymmetric alignment.}
\label{sec:FGM}
We begin by briefly reviewing FGM models~\cite{Shadmi:2011hs}.
The starting point in these models is mGMSB~\cite{dnns}.
Specifically, we will take $N$ sets 
of messengers  transforming as 5$+\bar5$ of SU(5),
coupled to a supersymmetry-breaking singlet 
$\langle X\rangle = M+F\theta^2$.
We use capital letters to denote the messenger fields, with
$5=T+D$ and $\bar 5=\bar T+\bar D$,  
where $T$ ($\bar T$) and $D$ ($\bar D$) are
fundamentals (anti-fundamentals) of $SU(3)$ and $SU(2)$ respectively. 
The SM gauge symmetry permits different couplings of the messengers
to SM fields, and these would generically give rise to flavor-dependent 
soft-terms~\cite{Dine:1996xk,Giudice:1997ni,Han:1998xy,Chacko:2001km,
Joaquim:2006uz,Joaquim:2006mn}\footnote{The model of~\cite{Chacko:2001km}
relies on an extra dimension in order to obtain MFV couplings.}. 
 
As in~\cite{Shadmi:2011hs}, we will assume that the SM fermion masses are
explained by a flavor symmetry. This symmetry then also controls
the messenger-matter couplings\footnote{In general, some messenger fields
may be charged under the flavor symmetry.}.
In the models we will consider, these coupling matrices will
be aligned with the SM Yukawa matrices, so that flavor constraints
are satisfied.
Naively, one would think that alignment can only be relevant
for high-scale supersymmetry breaking. This is indeed the case
in the original alignment models of~\cite{Nir:1993mx}. In these models,
a Froggatt-Nielsen flavor symmetry~\cite{Froggatt:1978nt} 
dictates the structure of the 
soft-term matrices at the supersymmetry-breaking scale.
As explained above however,
the non-universal parts of the soft terms in FGM models are
generated by superpotential matter-messenger couplings.
These {\sl supersymmetric} coupling matrices are the ones controlled
by the flavor symmetry at high scales, 
and their near-diagonal 
structure is inherited by the soft terms, which are generated
at much lower scales. We therefore refer to this type of alignment
as ``supersymmetric alignment''.

In addition to the flavor symmetry and R-parity, we
impose a $Z_3$ symmetry with charges given
in Table~\ref{tab:new_sym_charges}.
\begin{table}[h]
\centering
\renewcommand{\arraystretch}{1.25}
\begin{tabular}{|c|c|c|}
\hline
Superfield & $R$-parity & $Z_3$  \\
\hline\hline
$X$ & even & $1$  \\ \cline{1-3}
$D_1$ & even & $-1$  \\
$\bar{D}_1$ & even & $0$  \\
$D_2$ & even & $0$  \\
$\bar{D}_2$ & even & $-1$  \\\cline{1-3}
$T_I, \bar{T}_I, D_{I>2}, \bar{D}_{I >2}$ & even & $1$  \\ 
\cline{1-3}
$q,u^c,d^c,l,e^c$ & odd & $0$ \\
$H_U,H_D$ & even & $0$  \\
\hline
\end{tabular}
\caption{R-parity and $Z_3$  charges.}
\label{tab:new_sym_charges}
\end{table}
The following superpotential is then allowed by the symmetries,
 \beqa
\label{eq:superpot}
W = X \left( X^2 + T_I \bar{T}_I + D_I \bar{D}_I  \right)
&+&  H_U q Y_U u^c +  H_D q Y_D d^c +  H_D l Y_L e^c \nonumber\\ 
&+&  \bar D_1 q y_U u^c+  D_2 q y_D d^c + D_2 l  y_L e^c 
~,
\eeqa
where $I=1,\ldots,N$ runs over messenger pairs, $y_U$, $y_D$, $y_L$ 
are messenger-matter Yukawa matrices, 
$Y_U$, $Y_D$, $Y_L$ are the SM Yukawa matrices, 
and $q$, $u^c$, $d^c$, $l$, $e^c$ are the MSSM chiral multiplets.
We assume that the $\mu$-term(s) are
forbidden by some U(1) symmetry. 
Note that to have messenger couplings to both up quarks and down quarks
we need at least two sets of messengers~\cite{Chacko:2001km,Shadmi:2011hs}.
We also display here the term $X^3$, required in mGMSB
in order to generate the $X$ VEVs, and motivating our choice of a $Z_3$
symmetry. In the following, however, we will limit ourselves to treating
$X$ as a supersymmetry breaking spurion.

At this point, $D_2$ and $H_D$, as well as $\bar D_1$ and $H_U$, have identical
charges under all the symmetries, and therefore the following terms are 
allowed as well,
\beq\label{forbid}
X D_1 H_U + X H_D \bar D_2\ .
\eeq
However, we can set these couplings to zero without loss of generality.
Consider for concreteness
the $H_U$ and $\bar D_1$ couplings
\beq
{y_U}_{ij} \bar{D}_1 q_i u^c_j+{Y_U}_{ij}  H_U q_i u^c_j\ ,
\eeq
where $i,j$ are generation
indices.
Taking $H_U$ and $\bar D_1$ to have the same charges under the flavor symmetry,
we can define the combination of $\bar D_1$ and $H_U$ that couples
to $X$ to be the messenger (indeed, this is the massive eigenstate), 
and the orthogonal combination to be the Higgs. 
A similar redefinition can be done for $D_2$ and $H_d$. 
Thus~(\ref{eq:superpot}) is the most general superpotential and 
the entries ${y_U}_{ij}$ and ${Y_U}_{ij}$ 
are the same up to order-one coefficients\footnote{The running
between the UV scale and the messenger scale will introduce, of course,
some mixing between $H_U$ and $\bar D_1$, but the only effect
of this running is to modify the order-one coefficients of 
$y_U$ and $Y_U$.}. 
Since the only order-one entry of $Y_U$ is ${Y_U}_{33}$, the two
matrices $Y_U$ and $y_U$ are approximately diagonal,
and the soft terms inherit this structure.
Inter-generational mixings are thus suppressed by supersymmetric
alignment\footnote{
It is also possible to choose different charges for $H_U$ and
$\bar D_1$, (and similarly for $H_D$ and $D_2$) such that the terms~(\ref{forbid})
are either forbidden or very suppressed. In this case, $y_U$ and $Y_U$
will have different textures, and these can be chosen to
be compatible with flavor constraints~\cite{Shadmi:2011hs}.}.

\subsection{$A$-terms and scalar masses}
\label{sec:soft terms}
At leading order in $F/M^2$,
the messenger-matter couplings of~\eqref{eq:superpot} generate one-loop 
contributions to the $A$-terms, 
and two-loop contributions to the sfermion masses-squared.
We present full expressions for the soft terms in Appendix~A. 
In the case of interest, only the 3-3 entries of the coupling matrices 
are important and the soft terms (at the messenger scale) simplify to,
\begin{equation}
\begin{split}
A^U_{33}
&=
-\frac{1}{16\pi^2}
\left[  
3y^2_t 
+ y^2_b
\right]\frac{F}{M}
 \\
A^D_{33}
&=
-\frac{1}{16\pi^2}
\left[
3y^2_b
+y^2_t
\right]\frac{F}{M}
 \\
A^L_{33} 
&=
-\frac{3 y^2_\tau}{16\pi^2}
\frac{F}{M} \ ,
\end{split}
\end{equation}
where $Y_t\equiv {Y_U}_{33} $ and similarly for the remaining couplings,
and,

\begin{equation}\label{eq:masses3}
\begin{split} 
\tilde m^2_{H_U} 
&=
\frac1{128\pi^4} 
\left\{
-\frac{3}{2}Y^2_t(3y^2_t + y^2_b)
+N\left(\frac{3}{4}g_2^4 + \frac{3}{20}g_1^4\right) 
\right\}\, \left|\frac{F}{M}\right|^2
 \\ 
\tilde m^2_{H_D} 
&=
\frac1{128\pi^4} 
\left\{
-\frac{3}{2}Y^2_b(3y^2_b + y^2_t)
-\frac{3}{2}Y^2_\tau y^2_\tau
+N\left(\frac{3}{4}g_2^4 + \frac{3}{20}g_1^4 \right)
\right\}\, \left|\frac{F}{M}\right|^2 \\
(\tilde m^2_q)_{33} 
&=
\frac1{128\pi^4} 
\Bigg\{
\left(
y^2_t + 3y^2_b + 3 Y_{b}^{2}+ \frac{1}{2}y^2_\tau
-\frac{8}{3}g_3^2 - \frac{3}{2}g_2^2 - \frac{7}{30}g_1^2
\right) y^2_b
\\
&+\left(3y^2_t + 3Y_{t}^{2}
-\frac{8}{3}g_3^2 - \frac{3}{2}g_2^2 
- \frac{13}{30}g_1^2\right)y^2_t
+ Y_{b}y_{b}Y_{\tau}y_{\tau}
+N\left(\frac{4}{3}g_3^4 + \frac{3}{4}g_2^4 + \frac{1}{60}g_1^4\right)
\Bigg\}\, \left|\frac{F}{M}\right|^2
\\
(\tilde m^2_{u^c})_{33}
&=
\frac1{128\pi^4} 
\left\{
\left(
6y^2_t
+y^2_b
+Y^2_b
+6Y_{t}^{2}
-\frac{16}{3}g_3^2 - 3g_2^2 - \frac{13}{15}g_1^2
\right)y^2_t
-Y^2_ty^2_b\right.\\
&+\left.N\left(\frac{4}{3}g_3^4 + \frac{4}{15}g_1^4\right)
\right\}\,\left|\frac{F}{M}\right|^2 
 \\
(\tilde m^2_{d^c})_{33} 
&=
\frac1{128\pi^4} 
\left\{
\left(
6y^2_b
+y^2_\tau
+y^2_t
+Y^2_t
+6Y_{b}^{2}
-\frac{16}{3}g_3^2 - 3g_2^2 - \frac{7}{15}g_1^2 
\right) y^2_b
-y^2_t Y^2_b\right.\\
&+\left. 2Y_{b}y_{b}Y_{\tau}y_{\tau}+ N\left(\frac{4}{3}g_3^4  + \frac{1}{15}g_1^4\right)
\right\}\, \left|\frac{F}{M}\right|^2, 
 \\
(\tilde m^2_l)_{33} 
&=
\frac1{128\pi^4} 
\left\{
\left(
\frac{3}{2}y^2_b 
+ 2y^2_\tau
-\frac{3}{2}g_2^2 
-\frac{9}{10} g_1^2
\right) y^2_\tau
+\left(Y_{\tau}^{2}y_{\tau}^{2}+3 Y_{b}y_{b}Y_{\tau}y_{\tau}\right)
+N\left(\frac{3}{4}g_2^4 + \frac{3}{20}g_1^4\right)
\right\}\, \left|\frac{F}{M}\right|^2
 \\
(\tilde m^2_{e^c})_{33} 
&=
\frac1{128\pi^4} 
\left\{
\left(
3y^2_b +4 y^2_\tau\ -3g_2^2 -\frac{9}{5} g_1^2
\right) y^2_\tau
+\left(2 Y_{\tau}^{2}y_{\tau}^{2}+6 Y_{b}y_{b}Y_{\tau}y_{\tau}\right)
+\frac{3}{5}N g_1^4 
\right\}\,\left|\frac{F}{M}\right|^2 \,.
\end{split}
\end{equation}

If the messenger scale $M$ is below roughly $10^7\,\GeV$,
the one-loop $\mathcal{O}(F^4/M^6)$ contributions~\cite{Dine:1996xk} 
to the soft masses may be important. 
In the limit of third-generation dominance, 
these contributions are given by,
\begin{equation}
 \begin{split}
(\delta \tilde m^2_q)_{33} &= -\frac{1}{6}\frac{1}{16 \pi^2}\left(y_t^2+y_b^2
\right)
\frac{F^4}{M^6} \\
(\delta \tilde m^2_{u^c})_{33} &= -\frac{1}{3}\frac{1}{16 \pi^2}y_t^2
\frac{F^4}{M^6} \\
(\delta \tilde m^2_{d^c})_{33} &= -\frac{1}{3}\frac{1}{16 \pi^2}y_b^2
\frac{F^4}{M^6}
\\
(\delta \tilde m^2_{l})_{33} &= -\frac{1}{6}\frac{1}{16 \pi^2}y_\tau^2
\frac{F^4}{M^6} \\
(\delta \tilde m^2_{e^c})_{33} &= -\frac{1}{3}\frac{1}{16 \pi^2}y_\tau^2
\frac{F^4}{M^6} 
 \ .
\end{split}
\end{equation}
For completeness we also show the next-to-leading contribution in $F/M^2$
to the top $A$-term,
\beq
\delta A^U_{33} = -\frac{1}{16 \pi^2}y_t^2\frac{F^3}{M^5}\,.
\eeq

Comparing the new contributions to the mGMSB expressions, we see that 
the importance of the new contributions 
relative to the mGMSB expressions is maximal 
for 
the smallest possible
number of messengers. 
Thus we will take $N=1$ when the only new Yukawa coupling is $y_t$, 
and $N=2$ if $y_b$ and/or $y_\tau$ are present as well.

As is well known~\cite{Carena:2002es}, 
at one-loop, the Higgs mass is approximately given by,
\beq
m_h^2=m_Z^2 \cos^22\beta + \frac{3m_t^4}{4\pi^2v^2}\, 
\left[\log\frac{M_S^2}{m_t^2} 
+ \frac{X_t^2}{M_S^2}\,\left[1-\frac{X_t^2}{12M_S^2}\right] \,
\right]\,,
\eeq
where
$X_t=A_t -\mu \cot\beta$ is the LR stop mixing and
$M_S\equiv (m_{\tilde t_1}m_{\tilde t_2})^{1/2}$ is the average stop mass.
Clearly, the Higgs mass can be increased either by
increasing the average stop mass, so that the log term is large,
or by increasing the stop mixing, so that $X_t/M_S$ is large,
with the maximal $m_h^2$ obtained for $X_t/M_S$ of 
around 2.4~\cite{Haber:1996fp}.

Since our main objective is to obtain the correct Higgs mass with superpartners
within  LHC reach, we need a large $X_t/M_S$, and therefore
a large $A_t$. 
As can be seen from equations~(\ref{eq:masses3}), 
a non-zero $y_t$,
which generates $A_t$, also gives new contributions to the stop masses,
proportional to $y_t^4$.
These contributions are positive,
so that $M_S$ is increased as well.
Note that the new coupling also gives rise  to negative contributions
to the stop masses, proportional to $y_t^2 g^2$, but there is an accidental
cancellation between these and the positive $y_t^2 Y_t^2$
contribution. 
At low messenger scales however, the one-loop contributions are important, 
and since 
these are negative, one can obtain large $X_t/M_S$ with low $M_S$.

Thus, at low messenger scales, our models can give a heavy Higgs
together with light stops, while at high scales, a heavy Higgs necessitates
heavy stops. The messenger-scale masses of the remaining superpartner 
remain unchanged, and will only be modified by the running.

\section{Higgs mass and superpartner masses}
\label{sec:phenom}
\subsection{The Higgs and stop masses}
We first consider models with one set of messengers. 
With the $Z_3$ charges  of Table~\ref{tab:new_sym_charges}, 
only the $\bar D_1$ messenger couplings to matter are allowed.
Moreover, since we assume that $\bar D_1$ and $H_U$ have the same charge
under the flavor symmetry of the model,
the only significant
messenger coupling is $y_t\equiv (y_U)_{33}$.
We use SOFTSUSY~\cite{Allanach:2001kg} to calculate the Higgs mass
for different choices of the GMSB parameters and $y_t$.
Given the theoretical uncertainty in the calculation of    
the Higgs mass, it is interesting to study Higgs masses in the 
124--128 GeV window.

\begin{figure}[t] 
     \subfigure[]{
                \centering
                \includegraphics[width=0.455\textwidth]{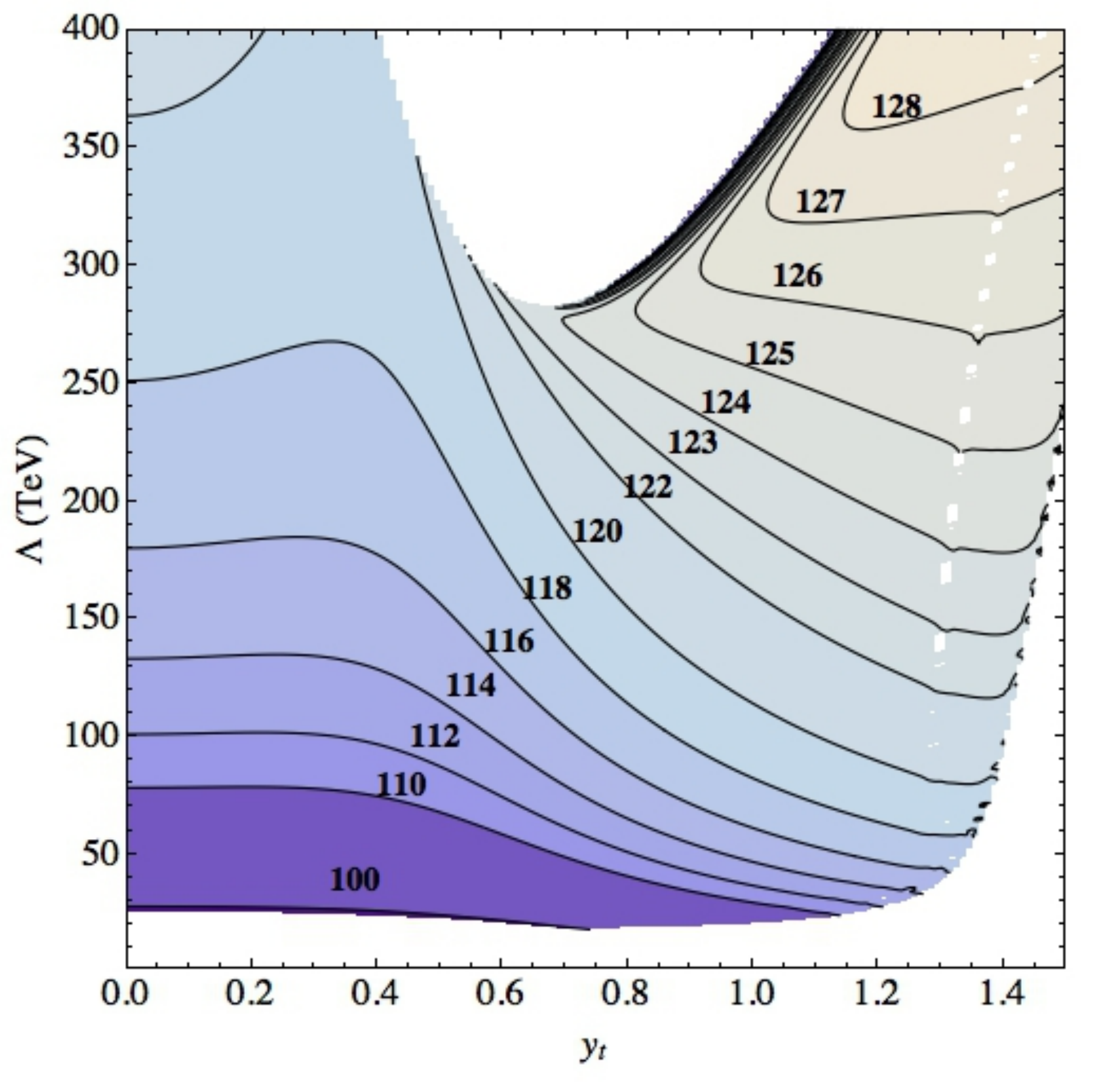}
		\label{fig:900a}
	}
     \subfigure[] {
                \centering
                \includegraphics[width=0.455\textwidth]{1b.pdf}
		\label{fig:900b}
	}
     \subfigure[] {
                \centering
                \includegraphics[width=0.455\textwidth]{1c.pdf}
		\label{fig:900c}
	}
\caption{The Higgs and stop masses for $N=1$, 
$M=900\,\mathrm{TeV}$, $\tan\beta=10$. 
Fig.~\ref{fig:900a} shows the Higgs mass for a wide range of $y_t$. 
The predictions of minimal gauge mediation can be read off from the 
line $y_t=0$. 
The white region is excluded because it leads to tachyonic stops (see text).
In Fig.~\ref{fig:900b}, we show Higgs mass (solid), 
heavy stop mass (dotted) and light stop mass (dashed) contour lines
in a smaller region of $y_t$.
In Fig.~\ref{fig:900c} we show Higgs  (solid), 
$\mu$ (dashed) and $x_t=|X_t/M_S|$ (dotted) contour lines
in the same region.
}
\label{fig:900}
\end{figure}

In~Fig.~\ref{fig:900} we show contours of the Higgs and stop masses
as a function of $\Lambda\equiv F/M$ and $y_t$, 
for a low messenger scale of
$M=900$~TeV, with $\tan\beta=10$.
For such a low messenger scale, 
the one-loop $\mathcal{O}(F^4/M^6)$ corrections are 
not necessarily negligible and have been taken into account. 
Fig.~\ref{fig:900a} shows the Higgs mass contours for a 
wide range of $y_t$. 
The white region for intermediate values of $y_t$ is excluded.
In this region,
the stops are either tachyonic or too light for successful electroweak
symmetry breaking. 
As explained above,
for these values of $y_t$, the negative one-loop contribution 
to the stop mass-squared is 
comparable to the positive contributions from pure GMSB.
As $y_t$ is increased, the $y_t^4$ contribution to the stop masses 
guarantees that the stops are non-tachyonic, but  because of the partial
cancellation between the 1-loop and 2-loop contributions,
the $A$-term becomes appreciable compared to the stop masses and the 
 resulting large mixing allows for a heavy Higgs. 
As $y_t$ is increased further, the stops become heavy relative
to the other superpartners. In this regime, the heavy stops 
and large A-terms both play a crucial role in making the Higgs heavy.

In Fig.~\ref{fig:900b},
we 
zoom in on the interesting 
range $y_t\sim1$, and show contours of the Higgs
mass together with the two stop masses,
with the remaining parameters being the same as in Fig.~\ref{fig:900a}.
In Fig.~\ref{fig:900c}, we also show contours of $\mu$ and the mixing parameter 
$x_t=|X_t/M_S|$.
As expected, the largest values of $x_t$ are obtained close to the 
excluded regions where one of the stops is relatively light. 
Thus, appreciable $A$-terms can be obtained without a large increase
 in the stop squared masses. 
Indeed, as can be seen in Fig.~\ref{fig:900b}, 
for these large values of $x_t$, the Higgs mass can be large
even for low $\Lambda$'s, such that the stops are light.
For $y_t\sim 1$ we can therefore find 
at least one stop below 2~TeV.

\begin{figure}[t] 
     \subfigure[] {
                \centering
                \includegraphics[width=0.45\textwidth]{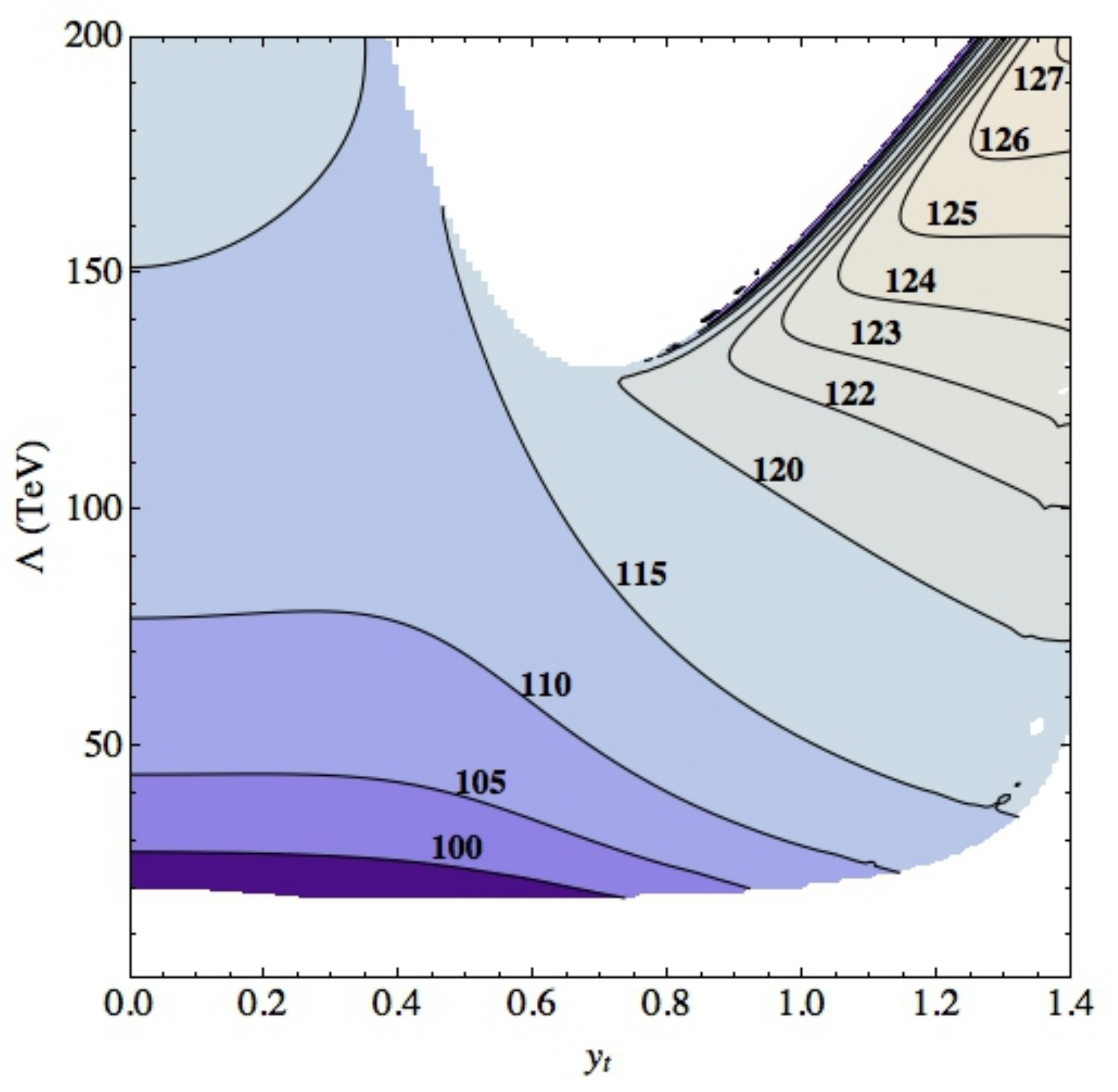}
                \label{fig:400a}
	}
     \subfigure[] {
                \centering
                \includegraphics[width=0.45\textwidth]{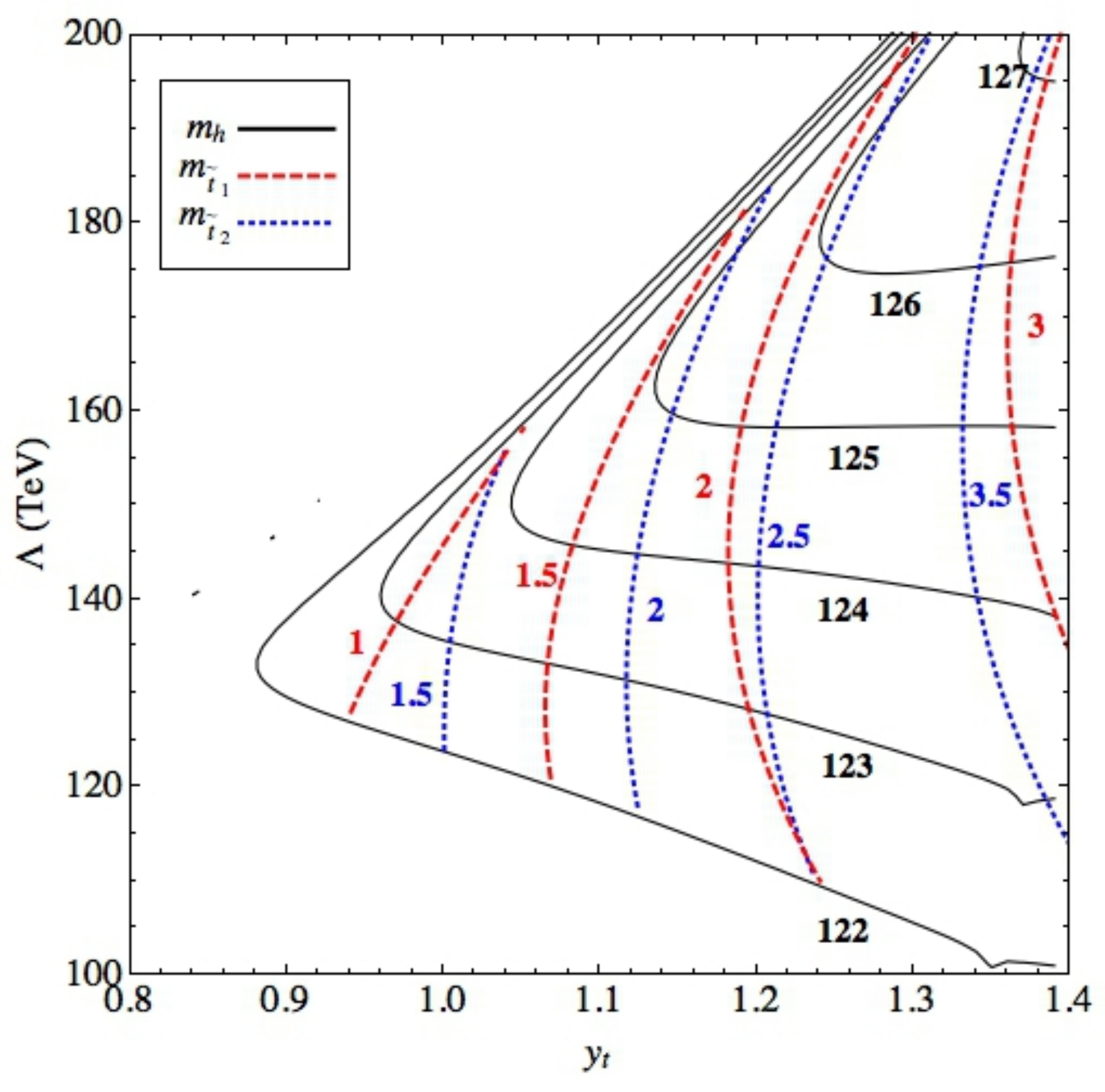}
                \label{fig:400b}
	}
\centering
\caption{Same plots as in Fig~\ref{fig:900a} and \ref{fig:900b} with
$M=400 \,\mathrm{TeV}$ and $\tan\beta=10$}
\label{fig:400}
\end{figure}

We can get even lighter stops by lowering the messenger scale, which allows
for a lower $\Lambda$. In Fig.~\ref{fig:400} we show the behavior of the models
for $M=400$~TeV. Note that since we only know the leading $F/M^2$ behavior
of the new contributions to the soft masses, we keep $F/M^2< 0.5$ (the
pure GMSB higher-order corrections  are known to be small~\cite{martin}).
Indeed, a Higgs mass of 125~GeV is obtained with both stops between
1.5 and 2~TeV, and stops below 1.5~TeV allow for Higgs masses above 124~GeV
(as mentioned above, one should bear in mind the uncertainty in our
Higgs mass calculation). The remaining squarks will be quite light
too in this region, and we will give a few example spectra to illustrate
this in the next Section.

\begin{figure}[t] 
     \subfigure[] {
                \centering
                \includegraphics[width=0.45\textwidth]{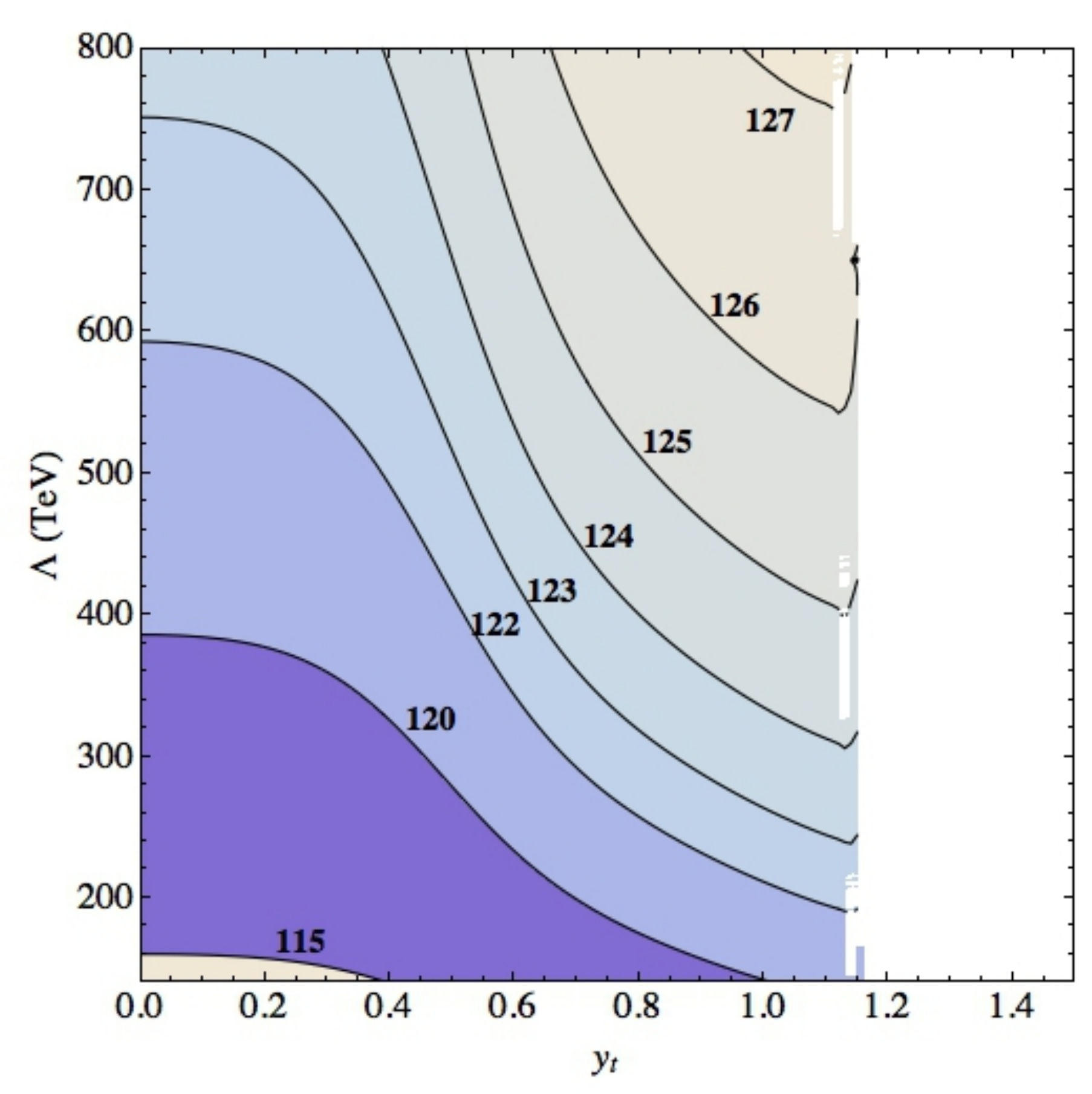}
                \label{fig:1e12a}
	}
     \subfigure[] {
                \centering
                \includegraphics[width=0.45\textwidth]{2b.pdf}
                \label{fig:1e12b}
	}
\centering
\caption{Same plots as in Fig~\ref{fig:900a} and \ref{fig:900b} with
$M=10^{12} \,\mathrm{GeV}$ and $\tan\beta=10$}
\label{fig:1e12}
\end{figure}

\begin{figure}[t] 
     \subfigure[] {
                \centering
                \includegraphics[width=0.455\textwidth]{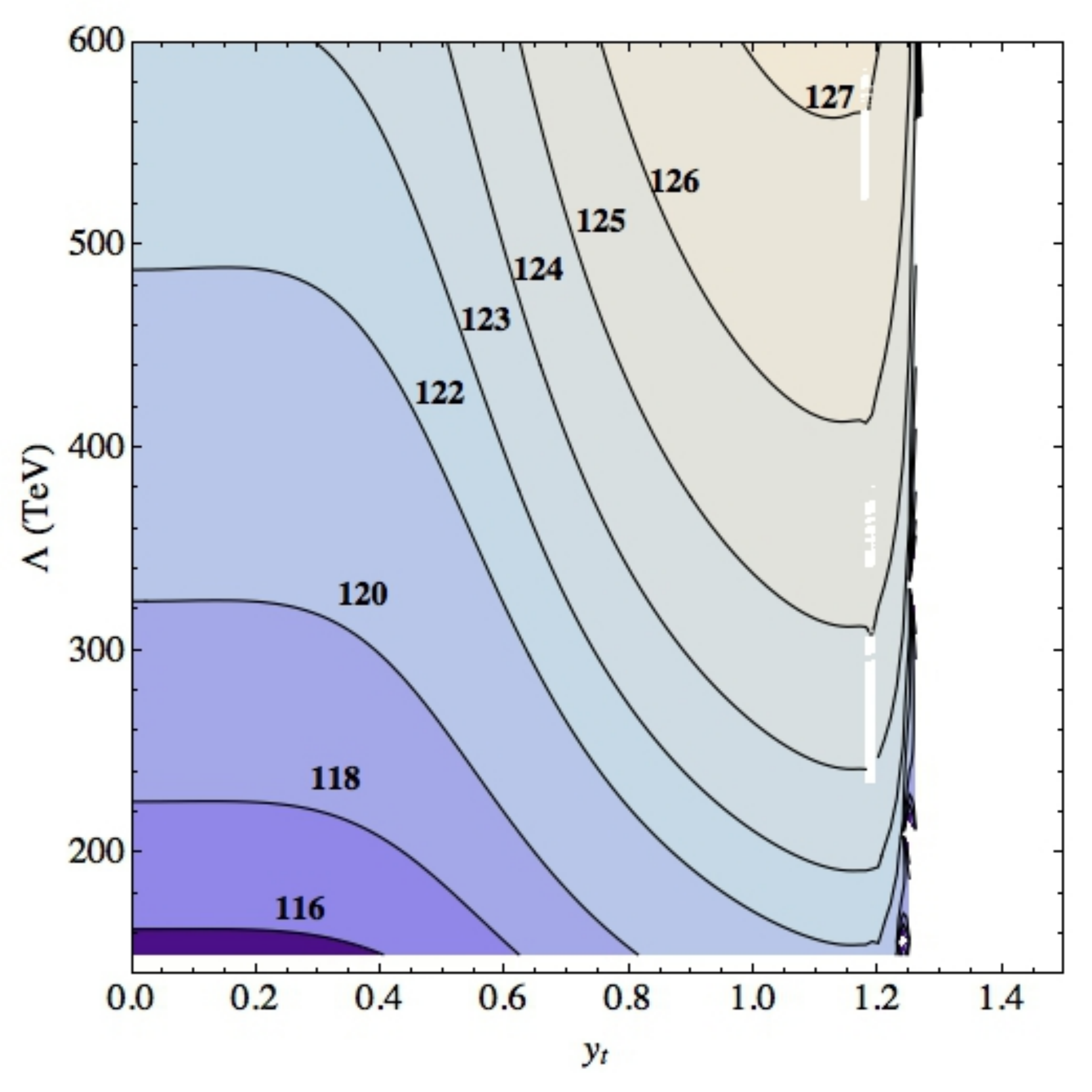}
                \label{fig:1e8a}
	}
     \subfigure[] {
                \centering
                \includegraphics[width=0.465\textwidth]{3b.pdf}
                \label{fig:1e8b}
	}
\centering
\caption{Same plots as in Fig~\ref{fig:900a} and and \ref{fig:900b} with  
$M=10^{8}\,\mathrm{GeV}$, $\tan\beta=20$. 
}\label{fig:1e8}
\end{figure}

For higher messengers scales, the behavior of the models is qualitatively
different.  
To demonstrate this,
we present similar plots for two other messenger scales, 
$M=10^{12}\,\GeV$ in Fig~\ref{fig:1e12} 
and 
$M=10^{8}\,\GeV$ in Fig~\ref{fig:1e8}. 
Clearly, the tachyonic stop region for moderate $y_t$ is absent 
for these high scales, since the negative one-loop contribution
is negligible. Thus a 125~GeV Higgs requires heavy stops, above 4~TeV.
It is 
interesting
to compare our results for high messenger scales 
with models of minimal gauge mediation. 
As is well known, with a lot of running, appreciable $A$-terms
can be generated in pure GMSB models. 
This is, however, not sufficient---as was shown in Ref.~\cite{Feng:2012rn}, 
even with a high messenger scale a heavy Higgs requires very heavy 
stops near 8-10~TeV. 
For example, with $M=7.9\cdot 10^{12} \,\GeV$ 
and $\tan\beta=10$, a Higgs mass of $125\,\GeV$ can be 
achieved if 
$\Lambda=1.3\cdot 10^6 \,\GeV$~\cite{Feng:2012rn}. 
With such a high value of $\Lambda$, one of the stops is the 
lightest squark and has a mass of 
$7.9\,\TeV$. 
In contrast,
if we choose the same messenger scale in the FGM model 
with $y_t=1.1$, a Higgs mass of $124.2\,\GeV$ can be achieved 
with 
$\Lambda=3.35\cdot10^5\,\GeV$. While the stops are still very heavy 
(around 5~TeV) due to the messenger Yukawa contributions, 
the remaining superpartners are significantly lighter, with the gluino and
right handed up and charm squarks around 2.3--2.4~TeV,
and a 461~GeV bino NLSP.

\begin{figure}[b] 
   \centering
   \includegraphics[width=0.4\textwidth]{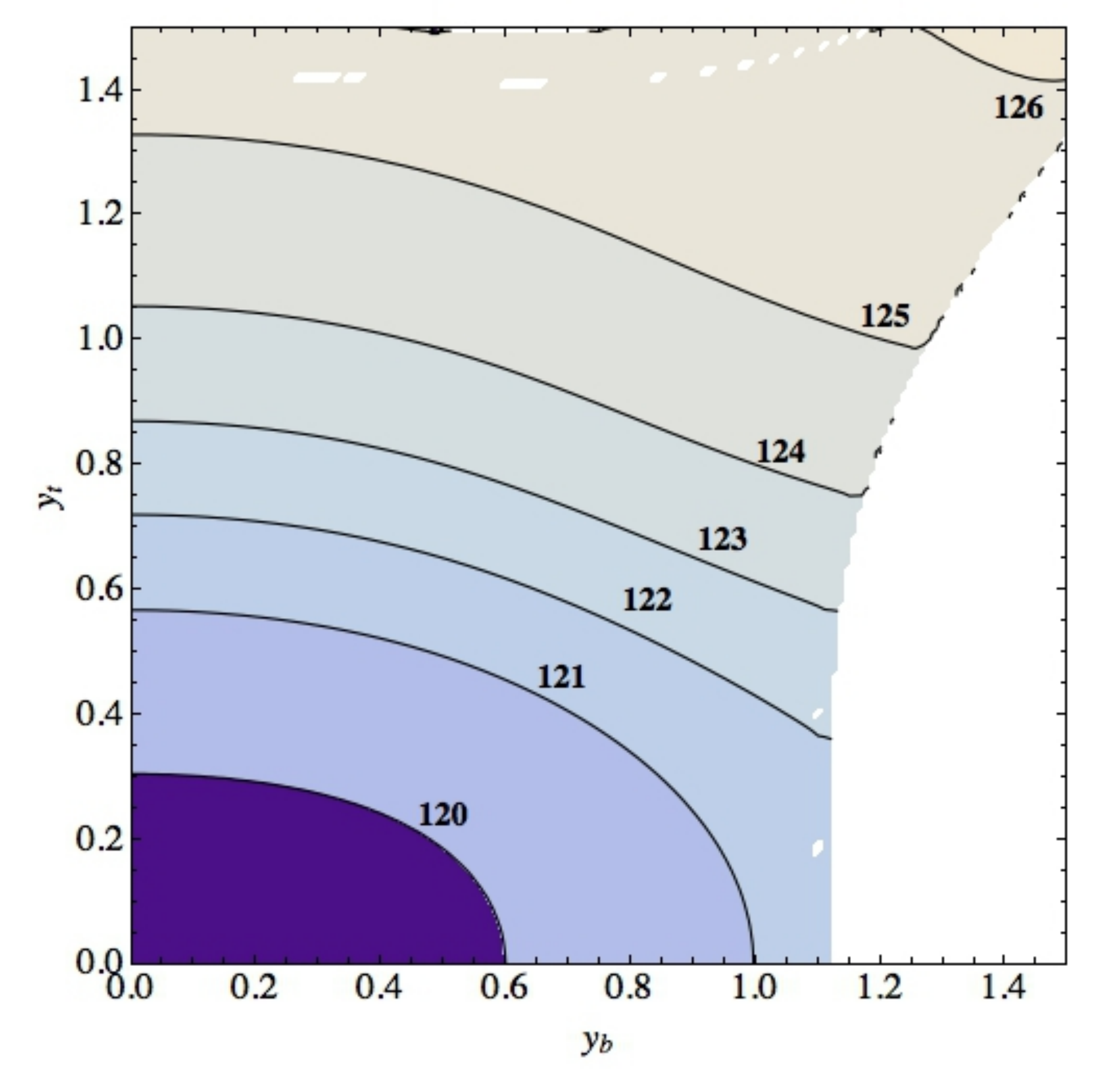}
   \caption{Higgs mass contours in the $y_t-y_b$ plane with 
$M=10^8\,\GeV$, $\Lambda=230\,\TeV$, and $\tan\beta=10$.}
   \label{fig:ytyb}
\end{figure}

While the models with only messenger-top Yukawas are the most economical ones, 
viable models with additional messenger-bottom and 
messenger-tau couplings may also be viable. 
As an example, in Fig.~\ref{fig:ytyb} we present Higgs mass 
contours in the $y_t-y_b$ plane 
for $N=2$, with  
$M=10^8\,\GeV$, $\Lambda=230\,\TeV$, and $\tan\beta=10$.
For large values of $y_b$, the sleptons become tachyonic, leading to
the white excluded region on the right.

Since we are turning on order-one superpotential couplings,
it is interesting to ask at what scales these become non-perturbative.
For example, if $N=1$ the high scale models with $M=10^{12}$~GeV 
remain perturbative
even above the GUT scale. For $M=10^8$~GeV, one loses perturbativity at
around $10^{12}-10^{13}$~GeV for $y_t\sim1$. Finally, with $M=900$~TeV,
the models stay perturbative up to $10^9-10^{10}$~GeV for $y_t$ above 1,
and for a smaller $y_t\sim0.9$ 
(as is the case with one of the lighter spectra we show
below. See Table~\ref{table:spectra}) up to scales of around 10$^{13}$~GeV.
In the case of $N=2$ the couplings remain perturbative for a few order 
of magnitude above the messenger scale.

\subsection{Superpartner spectra and LHC signatures}
To understand the phenomenology of our FGM models, we present in 
Table~\ref{table:spectra} 
\begin{table}
\begin{center}
$
\begin{array}{c|c|c|c|c|c|c|c|c}
{\rm
Parameter}&	 {\rm Spect.}\,\,\, 1		& {\rm Spect.}\,\,\, 2 
		& {\rm Spect.}\,\,\, 3	
	& {\rm Spect.}\,\,\, 4	& {\rm Spect.}\,\,\, 5 & {\rm Spect.}\,\,\, 6 & {\rm Spect.}\,\,\, 7 & {\rm Spect.}\,\,\, 8
		 \\ 
\hline
M_{\mathrm{mess}}	&	2\cdot10^{5}	&4\cdot10^{5}&9\cdot 10^5	&1\cdot 10^8	&1\cdot 10^8	&	1\cdot 10^{12}	
&1\cdot 10^{12}	   	&7.9\cdot10^{12}\\
\Lambda		
&	1.05\cdot10^5 	&1.65\cdot10^{5}	 &3.03\cdot 10^5	&3.08\cdot 10^5 &2.74\cdot 10^{5}&4.00\cdot 10^5 &3.55\cdot 10^5 &3.35\cdot10^{5}\\
\tan \beta		&10  &10&10 	&20	&20	&10	&10	 &10\\
y_t		&1.45	& 1.20&0.92	&1.19	&1.18&	1.13	&1.14		&1.10\\
\hline
\mu		&2606	&3165&4053	&6405	&5648&	7844	&7091		&6278\\
\hline
h_0		&124.9	&125.4&126.0	&125.0	&124.5&	125.0	&124.5		&124.2\\
A		&2686	&3281&4248	&6570	&5792&	8107	&7319		&6493\\
H_0		&2686	&3281&4250	&6567	&5791&	8105	&7319		&6493\\
H_{\pm}		&2687&3283&4249	&6571&5792	&	8107	&7320		&6494	\\
\hline
t_1		&1620	&1997&1795	&5698	&4986&	6243	&5634		&4899\\
t_2		&2050	&2315&2623	&7232	&6305&	7576	&6855		&5826\\
b_L		&1680	&2069&2615	&5654	&4948&	6195	&5591		&4864\\
b_R		&1119	&1683&2918	&2721	&2439&	3359	&3007	 	&2803\\
u_L,~c_L	&1179	&1780&3116	&2987	&2672&	3616	&3229		&3029\\
u_R,~c_R	&1096	&1668&2950	&2508	&2257&	2704	&2401		&2281	\\
d_L,~s_L	&1182	&1781&3117	&2987	&2673&	3617	&3230		&3030	\\
d_R,~s_R	&1133	&1698&2950	&2844	&2546&	3403	&3048		 &2839\\
e_L,~\mu_L	&305		&525	&1039	&569		&523&	578	&417			&620\\
e_R,~\mu_R	&356		&458&618		&1514	&1333&	2378	&2149			&1991\\
\tau_L		&244		&550&1038	&505		&462&	555	&390				&603\\
\tau_R	&392		&418&604		&1476	&1303&	2363	&2135			 &1979\\
\nu_e		&295		&519&1035	&555		&509&	562	&397		  	&606\\
\nu_\mu,~\nu_\tau	&295	 	&519 &1036&563	&516&	571	&408				&614	\\
\hline
\chi_1		&151	&233	&425		&426		&378		&552	&488			&461\\
\chi_2		&299	&457	&822		&826		&732&	1056	&935			&880\\
\chi_3		&2642	&3208&4107	&6573&5793	&	8010	&7239	 &6403\\
\chi_4		&2643	&3209&4108	&6573	&5793&	8010	&7240	   &6404	\\
\chi^{\pm}_1	&299	&458&823	&826	&733&	1056		&935	 	&880\\
\chi^{\pm}_2	&2643	&3210&4109	&6573&5794	&	8011	&7240	 &6404\\
g		&894	&1315&2240	&2251	&2024&	2832	&2540	 &2404\\
\end{array}
$
\caption{Model parameters, and resulting Higgs parameters and spectra 
for eight sample models. 
All mass parameters are given in GeV. }
\label{table:spectra}
\end{center}
\end{table}
complete superpartner spectra for several 
choices of the parameters at low, intermediate
and high messenger scales.
A detailed analysis of the experimental signatures is beyond the scope 
of this paper, and these spectra are only meant to illustrate the general 
features of the models. Thus for example, spectrum 1, which has a
light gluino and first generation squarks, is ruled out by jets-plus-missing-energy searches like~\cite{ATLAS:2012ona,Chatrchyan:2013lya} and 
possibly by specific GMSB searches~\cite{Aad:2012zza,Barnard:2012au}.
It is nonetheless useful to point out a few key
features.

First, large regions of the parameter space yield gluiono and squark masses that will hopefully be accessible in the 14~TeV LHC. Indeed, in many of the examples shown
in Table~\ref{table:spectra}, 
the gluino and some first generation squarks are below 2.5~TeV,
and sometimes considerably lighter.

Second,
even with a single messenger pair, for which mGMSB
models usually predict a neutralino NLSP, 
the NLSP in our models can be either a bino, 
a left-handed charged slepton, 
or a sneutrino, depending on the choice of parameters. 
This is due to the fact that the $U(1)_Y$ contributions to the RG evolution 
of sfermion masses contain a term proportional to 
$S\equiv\mathrm{Tr}[Y_j m_{\phi_j}^2]$, 
where the trace is taken over all the SM sfermions. 
Normally this contribution suppresses the right-handed sfermions masses, 
but in flavored GMSB the sign of $S$ changes for sufficiently large $y_t$. 
Furthermore, since for large $y_t$ the stop masses are much larger than
the remaining soft-terms, the effect of $S$ through the running is dramatic.
As a result, in some regions of the parameter space, the right-handed sleptons 
become heavy while the left-handed ones are relatively light.

While squarks are generically heavy in this class of models, 
the sleptons, charginos and neutralinos can be quite light, 
and may therefore be discovered even if produced directly.
This depends of course on
the details of the spectrum, and in particular on the identity
and lifetime of the NLSP.
Whether or not the  
NLSP
decays inside the detector depends on
the gravitino mass, which involves a lot
of uncertainty\footnote{
If $F_X$ is the dominant $F$-term, 
the gravitino mass in our models varies between
about 100~eV for $M=900$~TeV to a GeV for the high-scale models.
It is quite plausible however that there are much higher
$F$ terms in the supersymmetry-breaking sector, 
as is often the case
in calculable models, in which case  $F_X$ is generated through
several small couplings from a higher $F$-term. 
For large values of the gravitino mass, the NLSP would decay
outside the detector.
Heavy gravitinos from late NLSP decay could also supply
warm dark matter~\cite{Feng:2012rn,Feng:2008zza}.
}. 
In any case, the fact that the gravitino is the LSP in these models can
provide additional handles for their discovery, using either prompt
NLSP decays to the gravitino as in~\cite{Aad:2012zza,Barnard:2012au},
or the long lifetime of a charged NLSP. Thus for example, the current
bound on a long-lived charged slepton NLSP in mGMSB is 
300~GeV~\cite{Aad:2012pra}
and the model-independent bound on Drell-Yan produced right-handed sleptons
just somewhat below that. These bounds are likely to improve considerably
for long-lived charged left-handed sleptons at the 14~TeV LHC.

It would  also be interesting to study models with more messenger pairs.
Since the pure gauge contribution to the scalar squared masses
is proportional to $N$, 
we expect that for such models the cancellation
between the negative new contributions 
and pure gauge contributions will be less
dramatic than for the $N=1$ models, so that very light spectra 
would not occur. 
Such models would typically give rise to gluino masses above
squark masses, and to slepton or sneutrino NLSPs.

\section{Conclusions}
\label{sec:conclusions}
We have shown that gauge-mediated models with messenger-matter couplings 
can give rise to an acceptable Higgs mass, with 
colored superpartners within LHC reach.
The important new ingredient in these models is the messenger-scale
top $A$-term,
which gives rise to significant
stop mixing, and therefore enhances the Higgs mass. 
While a heavy Higgs can be obtained 
for a wide range of messenger scales, 
the details of the superpartner spectra may vary significantly. 
For low messenger scales, the one-loop $\mathcal{O}(F^4)$ negative 
contributions to the stop masses are 
important,
so that the stops are relatively light while the Higgs mass is raised largely 
due to the A-terms. 
For large messenger scales, the stops become heavy due to large positive 
contributions proportional to the messenger Yukawas. 
As a result, the Higgs mass is raised both due to the A-terms and to the large stop masses.

The messenger Yukawas often lead to another novel 
effect -- the $U(1)_Y$ contribution to sfermion RGE's 
may change sign so that left-handed sleptons become light. 
We see that the resulting spectra can be quite diverse with 
either a neutralino or slepton NLSP. We leave a detailed study
of the phenomenology of these models for future work. 
It would 
also be interesting to examine models with down-type messenger couplings, which may lead
to rather different phenomenology.
We further note that while we have concentrated
on models with only a single pair of messengers and a messenger-top
coupling, the results can be easily generalized to models with
several messenger pairs, with or without messenger down-type couplings.

The structure of the matter-messenger couplings can naturally
be the same as the structure of the usual Yukawas,
since the new couplings are obtained by replacing $H_U$ and/or $H_D$
by the messenger of the same gauge charge.
The models are therefore protected against flavor-violation
by {\sl supersymmetric alignment}. This can be simply realized 
in the context of flavor symmetries if the Higgs and corresponding
messenger have the same charges under the flavor symmetry.

The amount of tuning in our models
is related to the tuning of the new messenger-matter coupling $y_t$.
As can be seen in Table~\ref{table:spectra}, for spectrum~1,
which relies on a finely-dialed $y_t$ to obtain a light spectrum,
the $\mu$-term is 2.6~TeV, while less tuned choices of $y_t$ require
larger $\mu$-terms, around 6 or 8~TeV.
Generating an acceptable $\mu$-term in our models is an important
direction to pursue. If there is a successful mechanism for generating the
$\mu$-term, the tuning question would translate into the question
of how finely one needs to tune the coupling $y_t$. It is probably
far from trivial to find a successful mechanism for generating
the $\mu$-term, and even if it is found, the required tuning
of $y_t$ is likely to be significant. Still, since $y_t$ is
a superpotential coupling, the tuning involved is
qualitatively different from the tuning of the Higgs mass
in the standard model.

\section*{Note added}
After posting the original version of this article on the archive, M.~Ibe
drew our attention to~\cite{Evans:2011bea,Evans:2012hg}, which use the same messenger-top coupling
in order to raise the Higgs mass. There is thus some overlap between
our paper and~\cite{Evans:2011bea,Evans:2012hg} as concerns the implications 
of a heavy Higgs. 
The origin of the messenger-matter
coupling is different however in~\cite{Evans:2011bea,Evans:2012hg}, with the result that these
models are MFV. Also, while we calculate the new contributions 
of the full 3-generation coupling matrices, ~\cite{Evans:2011bea,Evans:2012hg} 
only consider
third generation couplings. 
While completing this revised version, Ref.~\cite{Evans:2013kxa} appeared,
which surveys different messenger-matter couplings.
Our results for the soft masses now agree with~\cite{Evans:2011bea,Evans:2012hg,
Evans:2013kxa},
but our approach to the calculation differs from theirs.

\section*{Acknowledgments}
We thank Nathaniel Craig, Daniel Feldman, Jonathan Feng,
Gilad Perez, Arvind Rajaraman, David Shih, and Carlos Wagner for useful
discussions. We are grateful to M.~Ibe for drawing our attention
to~\cite{Evans:2011bea,Evans:2012hg}, and for discussions on the calculation
of the soft terms. 
The research of M.~Abdullah was supported by Kuwait University.
The research of Y.~Shadmi and I.~Galon was supported by
the Israel Science Foundation (ISF) under grant No.~1367/11, 
and by the United States-Israel
Binational Science Foundation (BSF) under grant No.~2010221.
Y.~Shirman was supported by the National Science Foundation under 
grant PHY-0970173. Y.~Shirman is grateful to the Technion for 
hospitality during the early stages of this project.
Y.~Shadmi 
acknowledges the hospitality of UC Irvine, and of the 
Aspen Center for Physics, which is supported by the National Science 
Foundation Grant No. PHY-1066293.

\appendix
\section{Derivation of the soft terms}
In this Appendix, we describe the calculation of the soft terms.
As pointed out in~\cite{Giudice:1997ni}, these can be extracted
from the wave function renormalizations of the light fields,
treating the heavy threshold as a spurion. 
The main advantage of this method is that the running of the
wave function renormalizations, as well as of the various couplings,
is only needed at one-loop.
The method of~\cite{Giudice:1997ni} was used in~\cite{Chacko:2001km}
to obtain the soft terms in models with messenger-matter couplings.
However, the analysis of~\cite{Chacko:2001km}, as well as the analysis
in an earlier version of this article, did not treat the matter-messenger
mixing correctly. In~\ref{analcontapp} we clarify this issue by discussing a
simple toy model.
In~\ref{multiple}, we generalize the results to models with multiple fields
and couplings. Our final results are given in~\ref{final}.
 As a double check of our derivation,  we explicitly calculate 
the relevant contributions,
namely the mixed $y^2 Y^2$ terms, in~\ref{explicitcal}.

\subsection{Analytic continuation in the presence of mixing}
\label{analcontapp}
To discuss the calculation of the soft terms in the presence of 
messenger-SM mixings,
we first consider a simple toy model, with the superpotential
 \beq 
W =
 X \bar D  D + (y^0 D + Y^0 H) l e\,.
\eeq
Here  $X= M+F\theta^2$ is the SUSY breaking spurion, 
$\bar D$, $D$, $H$, $l$ and $e$ are singlet fields,
and we use  the superscript $^0$ to denote the superpotential couplings
$y^0$ and $Y^0$ in order to distinguish them from the running couplings.

Our analysis closely follows~\cite{Chacko:2001km}, which applied
the method of~\cite{Giudice:1997ni} to models with multiple couplings,
in which one cannot
integrate the one-loop RGEs to obtain closed-form expressions for the
wave function renormalizations and couplings.
The necessary ingredients in the calculation are the RGE's for the
various couplings, and the boundary conditions for these couplings.
In the absence of mixing between the messengers and SM fields,
there is a clear distinction between the messengers and light fields.
In our toy model however, $H$ and $D$ mix, and as a result, there is
some ambiguity in the identification of the messenger and Higgs couplings.
The key in the calculation is therefore the correct
matching of the high-energy and low-energy theories,
which we will perform by 
identifying the {\sl physical} heavy and light combinations of messenger and
Higgs fields, and by demanding that the physical coupling of the light
combination is continuous across the threshold.

Let us first recall the main results of~\cite{Giudice:1997ni}. 
At leading order in $F/M^2$, 
the soft mass of the light field $f$ can 
be extracted from its wave function renormalization $Z_f$,
\beq\label{analcont}
m^2_f(\mu) = -\frac14 \frac{\partial^2 \ln Z_f(\mu)}{\partial \ln M^2}\, 
\frac{F^2}{M^2}\,,
\eeq
at $\mu\leq M$.
This relies on the fact that, at this order, 
the only threshold dependence enters through the one-loop running
of $Z_f$, and therefore one can obtain
the soft masses by promoting $M$ to a superfield.
Specifically, as argued in~\cite{Giudice:1997ni} based on symmetries and holomorphy,
$Z_f(M)\to Z_f(\sqrt{X^\dagger X})$.
Note that, since the theory is defined at a scale $\Lambda$ above $M$,
the derivatives with respect to $M$ are taken while holding the
physical couplings at $\Lambda$ fixed. It is therefore natural to choose
a canonical K\"ahler  potential at $\Lambda$, and to hold
the superpotential couplings 
fixed while taking derivatives
with respect to $M$.

It will be convenient to rewrite our model by defining,
$\phi_1\equiv D$, $\phi_2\equiv H$, $y_1^0\equiv y^0$ and $y_2^0\equiv Y^0$.
The high energy theory is then defined at the cutoff $\Lambda$,  
by the superpotential
\beq 
W =
 X \bar D  \phi_1+ y^0_i \phi_i l e  
\eeq
where $i=1,2$. As noted above, we take the K\"ahler potential to 
be canonical at $\Lambda$.

At any scale $\mu$ below $\Lambda$ and above the 
messenger scale, the renormalized
fields are
\beq
\phi_r(\mu)_i\equiv Z^{1/2}(\mu)_{ij}\, \phi_j\,~~
l_r(\mu)\equiv Z_l^{1/2}(\mu) \, l\,,~~e_r(\mu)\equiv Z_e^{1/2}(\mu) \, e\,,
\eeq
Here $Z^{1/2}$ is the square-root of the two-by-two wave-function 
renormalization matrix $Z$.
The running couplings are given by
\beq\label{runningcoupling}
y_i(\mu)=   Z_l^{-1/2}(\mu)\, Z_e^{-1/2}(\mu)\, Z^{-1/2}(\mu)_{ji}\, y_j^0 \,.
\eeq
 Note that $Z$ is a real superfield. 
At one loop,
$Z$ runs according to
\beq\label{matrixrun}
\frac{dZ}{dt}=Z^{1/2}\gamma Z^{1/2} \,,
\eeq
where $\gamma$ is the two-by-two matrix of anomalous dimensions.

At the messenger scale 
$\mu_X$, we have a heavy combination $\tilde\phi_{r1}$ and (the orthogonal) 
light combination $\tilde\phi_{r2}$,
\beqa
\tilde\phi_{r1}&=&\left[Z^{-1/2}(\mu_X)_{11}\phi_{r1} + Z^{-1/2}(\mu_X)_{12}\phi_{r2}
\right]/C(\mu_X)\\
\tilde\phi_{r2}&=&\left[Z^{-1/2}(\mu_X)_{11}\phi_{r2} -
Z^{-1/2}(\mu_X)_{21}\phi_{r1} 
\right]/C(\mu_X)
\eeqa
where
\beq
C(\mu_X) = \sqrt{(Z^{-1/2}(\mu_X)_{11})^2 + |Z^{-1/2}(\mu_X)_{12}|^2}
\eeq
and where we used
\beq
(Z^{-1/2 *})_{ij}= (Z^{-1/2 })_{ji}
\eeq

We can now find the physical couplings at the threshold.
The physical coupling of the heavy combination $\tilde\phi_{r1}$
to $l$ and $e$ is,
\beq\label{physcoupl1o}
\tilde y_1(\mu_X) = Z_l^{-1/2}(\mu_X)\, Z_e^{-1/2}(\mu_X)\, 
C(\mu_X) \left[ y_1^0  + 
\frac{y_2^0  Z^{-1/2}_{21}(\mu_X)
\left(Z^{-1/2}_{11}(\mu_X)+Z^{-1/2}_{22}(\mu_X)\right)}{ C(\mu_X)^2} 
\right]\,,
\eeq 
and the physical coupling of the light combination $\tilde\phi_{r2}$
to $l$ and $e$ is,
\beq\label{physcoupl2o}
\tilde y_2(\mu_X) = Z_l^{-1/2}(\mu_X)\, Z_e^{-1/2}(\mu_X)\, 
\frac{
 Z^{-1/2}(\mu_X)_{11} Z^{-1/2}(\mu_X)_{22} -
\vert Z^{-1/2}(\mu_X)_{12}\vert^2 }
{\sqrt{|Z^{-1/2}(\mu_X)_{11}|^2 + |Z^{-1/2}(\mu_X)_{12}|^2}}\,
y^0_2 \,.
\eeq 
The physical messenger scale is
\beq\label{mux}
\mu_X^2=\left(|Z^{-1/2}(\mu_X)_{11}|^2 + |Z^{-1/2}(\mu_X)_{12}|^2\right)\,  
X^\dagger X\,.
\eeq
To leading order, we can replace $\mu_X^2$ by
$X^\dagger X$, since the difference between the two gives a  3-loop correction
to the soft masses  (see also~\cite{ArkaniHamed:1998kj}).
In the following we will therefore set $\mu_X=M$.
Furthermore, the expression for the soft masses~(\ref{analcont})
involves the running of $Z$ and the couplings at one-loop only.
Thus we only need to match the couplings at the threshold at one-loop,
and at this order the coupling of the heavy combination is
\beq\label{physcoupl1}
\tilde y_1(M) = Z_l^{-1/2}(M)\, Z_e^{-1/2}(M) \, Z^{-1/2}(M)_{11}\, 
\left(y_1^0 +
2  Z^{-1/2}(M)_{21} \, y_2^0 \right)\,,
\eeq
while the coupling of the light combination is,
\beq\label{physcoupl}
\tilde y_2(M) = Z_l^{-1/2}(M)\, Z_e^{-1/2}(M)\,Z^{-1/2}(M)_{22} \, 
y^0_2\ . 
\eeq 
Equations~(\ref{physcoupl1}) and~(\ref{physcoupl}) are the key results
of the preceding analysis, and lead to the main difference between our results
and the results of~\cite{Chacko:2001km}. 
The point is that 
these couplings do not coincide with the running couplings
$y_i(M)$ of eqn.~(\ref{runningcoupling}).
In particular, the coupling of the light combination at the threshold,
$\tilde y_2(M)$, does not involve either $y_1^0$ or
the mixed anomalous dimension $\gamma_{12}$ since at one loop 
$(Z^{-1/2})_{22}$ only depends on $\gamma_{22}$. 
Consequently, as we will see below, the soft mass of the light combination 
$H$ does not depend on the mixed anomalous dimension $\gamma_{12}$.
This is precisely what one would expect intuitively\footnote{In fact,
in the earlier version of this article, this intuition motivated us to
ignore the contributions of $\gamma_{12}$ in the soft masses. This
is indeed correct for $H$, but not for the other SM fields.}.
On the other hand,
the coupling of the heavy combination~(\ref{physcoupl1}) involves both
$y_1^0$ and $y_2^0$, with the latter multiplied by the
mixed anomalous dimension $\gamma_{12}$.  
However,  this contribution
 appears with a factor of 2 compared to the analogous term in
the running coupling $y_1(M)$.

The two conclusions of the above discussion, namely,
the absence of $\gamma_{12}$ in the coupling of the light combination,
and the factor of 2 multiplying $\gamma_{12}$ in the coupling of the 
heavy combination, only involve the fields $H$ and $D$, and are not affected
by the structure of the couplings to the remaining fields $l$ and $e$.
These conclusions therefore carry over trivially to the full 3-generation
model. In other words, we only need to integrate out the heavy field once,
and at one-loop, this procedure only involves the wave-function 
renormalizations 
of $H$ and $D$.

We can now turn to the low energy theory. 
Clearly, this theory can be 
written in terms of the fields $l$, $e$, and $H$.
Its coupling is defined by matching to the high scale theory at the threshold.
That is, we require that the running coupling in the low-energy theory,
$Y(\mu)$  match the physical coupling of the light combination 
at the threshold $M$ at one-loop,
\beq
Y(M) = \tilde y_2(M)
\eeq
with $\tilde y_2(M)$ given by eqn.~(\ref{physcoupl}).
The low-energy theory is therefore defined by
\beq
W= Y^0 H l e
\eeq
with $Z_l(M)$, $Z_e(M)$ and $Z_H(M)$ continuous across the threshold.
More precisely, for the latter\footnote{Note that at one loop,
$(Z_{22}^{-1/2})^2= Z_{22}^{-1}$.},
\beq
Z_H(M)= Z_{22}(M)\,.
\eeq
Thus, both the wave-function renormalization and the physical coupling
of the light combination are continuous across the threshold as one would 
expect, but, as noted above, the coupling of this combination 
is different from the running coupling $y_2(M)$.

Note that,
since we only have a single combination of $\phi_1$ and $\phi_2$ 
in the low-energy theory,
we have reverted to the original notation  and replaced $y_2^0$ by
$Y^0$. The running coupling below $M$ is therefore
\beq
Y(t) = \frac{Z_l^{-1/2}(t)}{Z_l^{-1/2}(M)} \,
\frac{Z_e^{-1/2}(t)}{Z_e^{-1/2}(M)}\,
\frac{Z_H^{-1/2}(t)}{Z_H^{-1/2}(M)}\, \tilde y_2(M) \,.
\eeq

\subsubsection{slepton mass}
Let us use this to calculate the $l$ mass following~\cite{Chacko:2001km}.
For $\mu<M$,
\beq\label{zL}
\ln Z_l(\mu)= -\int_{\ln M}^{\ln\Lambda} \gamma_l^>(t^\prime) \, dt^\prime -
\int_t^{\ln M} \gamma_l^< (t^\prime) \,dt^\prime \,,
\eeq
with $t=\ln\mu$. We use the superscript $>$ ($<$) to denote the theory
above (below) $M$.

We have
\beq
\frac{\partial}{\partial \ln M} \ln Z_l(\mu)= \gamma_l^>(M) -\gamma_l^<(M)
-\int_t^{\ln M} \frac{\partial}{\partial\ln M}\gamma_l^<(t^\prime) \, dt^\prime  \,,
\eeq
and
\beq\label{lmass1}
\frac{\partial^2}{\partial {\ln M}^2} \ln Z_l(M)= 
\frac{\partial}{\partial\ln M}
(\gamma_l^>( M) -\gamma_l^<( M))
-
\frac{\partial}{\partial\ln M}\gamma_l^<(t)\vert_{t=\ln M} \,.
\eeq
The jump in the $l$ anomalous dimension is given by the contribution
of the heavy field to $\gamma_l$. Therefore,
\beq
\gamma_l^>( M) -\gamma_l^<( M) =-\frac2{16\pi^2} \,\vert\tilde y_1( M)\vert^2.
\eeq
The $l$ anomalous dimension at scales below $M$ is given by
\beq
\gamma_l^<(t) =-\frac2{16\pi^2} \,\vert Y(t)\vert^2\,,
\eeq
with
\beq
\vert Y(t)\vert^2=\frac{Z_l( M)}{Z_l(\mu)} \frac{Z_e( M)}{Z_e(\mu)} 
 \frac{Z_h( M)}{Z_H(\mu)} Y(M)= Z_l^{-1}(\mu) Z_e^{-1}(\mu)  Z_H^{-1}(\mu) 
\,\vert y_2^0\vert^2\,.
\eeq
Thus,
\beq\label{lformula}
m_l^2(M)=-\frac14\,
\left|\frac{F}{M}\right|^2
\frac{\partial^2}{\partial {\ln M}^2} \ln Z_l(M) =
\frac14\,
\left|\frac{F}{M}\right|^2
 \frac2{16\pi^2}\, \left[
\frac{\partial}{\partial\ln M} \vert\tilde y_1( M)\vert^2 +
 \frac{\partial}{\partial\ln M}  \vert Y(t)\vert^2\vert_{t=\ln M} 
\right]\,.
\eeq
The derivatives in the expression above involve of course the beta
functions of the two couplings, which in turn are combinations
of the various anomalous dimensions. 
The derivative of the first term can be obtained from eqn.~(\ref{physcoupl1}),
and, as explained above, involves {\sl double} the usual contribution
of $\gamma_{12}$.

In contrast, the derivative of the second term does not contain $\gamma_{12}$
at all, since $\gamma_{12}$ cannot appear in the theory below $M$,
and does not appear in $Y(M)$ as we saw above. 
To obtain the second term of eqn.~(\ref{lformula}), we can use first
eqn.~(\ref{zL}) which gives at one loop,
\beq
\frac{\partial}{\partial\ln M} \ln Z_l(\mu) = \Delta \gamma_l(M)=
\gamma_l^>(M) -\gamma_l^<(M)
\eeq
so that
\beq
\frac{\partial}{\partial\ln M} Z_l^{-1}(\mu)= -\Delta \gamma_l(M)\,.
\eeq
We also need the analogous expression for $Z_H$,
\beq\label{zoneloop}
\ln Z_H(\mu) = \ln Z_{22}( M) - \int_t^{\ln M} \gamma_H \, dt^\prime 
\eeq
so at one-loop
\beq
\frac{\partial}{\partial\ln M} Z_H(\mu) =
\gamma_{22}( M) -\gamma_H( M) =0 \,.
\eeq

Plugging these in eqn.~(\ref{lformula}), we have,
\beq\label{finall}
m_l^2(M) = \frac14\, \frac2{16\pi^2}\,\left[
(\gamma_l^>+\gamma_e^>+\gamma_{11}^>) y^2 
+2\gamma_{12} y Y
-[\Delta\gamma_l +\Delta\gamma_e] Y^2
\right] 
\left|\frac{F}{M}\right|^2
\eeq
with everything evaluated at the scale $M$.
Substituting in the values of the anomalous dimensions
one gets
\beq
m_l^2 =  \frac1{(4\pi)^4}\, (4 y^4 + 2 y^2 Y^2)
\left|\frac{F}{M}\right|^2
\,,
\eeq
or for a simplified model where all fields are singlets
\beq
\label{eq:lsoft_simplified}
m_l^2\bigg|_{\text{simplified}} =  \frac1{(4\pi)^4}\, (3 y^4 + 2 y^2 Y^2)
\left|\frac{F}{M}\right|^2
\,,
\eeq

Alternatively, we can rewrite eqn.~(\ref{finall}) 
in a way that is more similar
to the expression of~\cite{Chacko:2001km},
\beq\label{mlcp}
\tilde m^2(M) = -\frac{1}{4}  
\left(
\frac{d\Delta \gamma}{dy} \left[\beta_y^<\right]_{\gamma_{12}=0}
-\gamma_{12}\frac{d\Delta \gamma}{dy} Y 
- \frac{d\gamma^<}{dY}\left[\Delta\beta_Y\right]_{\gamma_{12}=0}
\right)\,
\left|\frac{F}{M}\right|^2\,.
\eeq
Here the various $\beta$'s and anomalous dimensions are the standard
ones, and $[]_{\gamma_{12}=0}$ indicates that $\gamma_{12}$ should be set
to zero in the expression for the relevant $\beta$.
One could in principle denote the couplings collectively by $\lambda$,
as in~\cite{Chacko:2001km}, but $\Delta\gamma$ only depends on the messenger couplings 
$y$, whereas $\gamma^<$ only depends on the Higgs couplings $Y$.

\subsubsection{Higgs mass}
To calculate the Higgs mass we again need to take two derivatives of
\beq\label{zH}
\ln Z_H(\mu)= \ln Z_{22}(M) -
\int_t^{\ln M} \gamma_H^<(t^\prime) \, dt^\prime\,. 
\eeq
Since there is no jump in the anomalous dimension of $H$, one could immediately
start with the analog of eqn.~(\ref{lmass1}) and set $\Delta\gamma_H=0$.
We can also derive this result more carefully.
Writing
\beq
Z_{22}(M)= 1-
 \int_{\ln M}^{\ln\Lambda} \left(Z^{1/2} \gamma  Z^{1/2} \right)_{22} dt
\eeq
we find (dropping 3-loop terms)
\beq
\frac{\partial^2}{\partial{\ln M}^2} \ln Z_H(M)
=
\frac{\partial}{\partial{\ln M}} \gamma_{22}(M) + \vert \gamma_{12}\vert^2(M)\,.
\eeq
Then
\beq\label{step}
\frac{\partial^2}{\partial{\ln M}^2} \ln Z_H(M)\vert_{\mu=M}
 = 
\frac{\partial}{\partial\ln M}
(\gamma_{22}( M) -\gamma_H( M)) + \vert \gamma_{12}\vert^2
-
\frac{\partial}{\partial\ln M}\gamma_H^<(t)\vert_{t=\ln M} \,.
\eeq
Note that $\gamma_{22}(M)$ and $\gamma_H(M)$ differ at one-loop:
\beq
\gamma_{22}(M)=\frac{-2}{16\pi^2} \, \vert y_2(M)\vert^2=
\frac{-2}{16\pi^2} \, \left[
\vert y_2^0\vert^2 Z^{-1}_{22}(M) \, Z^{-1}_l(M) \, Z^{-1}_e(M)
+ \left( y_1^0 y_2^{0*}  Z^{-1/2}_{12}(M) +cc\right)
\right]\,,
\eeq
whereas
\beq
\gamma_H(M)=\frac{-2}{16\pi^2} \, \vert \tilde y_2(M)\vert^2=
\frac{-2}{16\pi^2} \,
\vert y_2^0\vert^2 Z^{-1}_{22}(M) \, Z^{-1}_l(M) \, Z^{-1}_e(M)\,.
\eeq
Therefore
\beq
\frac{\partial}{\partial\ln M}
(\gamma_{22}( M) -\gamma_H( M)) = \frac{-2}{16\pi^2}\, 
\frac{\partial}{\partial\ln M} 
\left( y_1^0 y_2^{0*}  Z^{-1/2}_{12}(M) +cc\right)=
-\vert \gamma_{12}(M)\vert^2\,,
\eeq
which precisely cancels the third term in~(\ref{step}).
We are then left with
\beq\label{naive1}
\frac{\partial^2}{\partial{\ln M}^2} \ln Z_H(M)
\Big\vert_{\mu=M} = 
-
\frac{\partial}{\partial\ln M}\gamma_H(t)\vert_{t=\ln M} \,.
\eeq
In this case, the soft mass only depends on the anomalous dimension
in the low-energy theory, and therefore does not involve the mixed
anomalous dimension $\gamma_{12}$ as explained in the previous section.

Using the results of the last section we find,
\beq\label{hfinal}
m_H^2 = \frac14\, \frac2{16\pi^2}\, 
[\Delta\gamma_l +\Delta\gamma_e] \vert Y\vert^2
\left|\frac{F}{M}\right|^2
=
-\frac1{(4\pi)^4}\, (3 \vert y\vert^2 \vert Y\vert^2)
\left|\frac{F}{M}\right|^2
\,,
\eeq
or for a simplified model where all fields are singlets
\beq
\label{eq:hsoft_simplified}
m_H^2 = 
- \frac1{(4\pi)^4}\, (2 \vert y\vert^2 \vert Y\vert^2)
\left|\frac{F}{M}\right|^2
\,.
\eeq
Again, we can rewrite this in analogy with eq.(\ref{mlcp}),
\beq
\tilde m_H^2(M) = \frac{1}{4}\,  
\frac{d\gamma_H^<}{dY}\left[\Delta\beta_Y\right]_{\gamma_{12}=0}
\,
\left|\frac{F}{M}\right|^2\,,
\eeq
where we used $\Delta\gamma_H=0$.

\subsection{Multiple couplings}\label{multiple}
We can now generalize these results to models with multiple fields
and couplings. Specifically, we will take the superpotential to be
 \beq 
W =
 X \bar D  D + (y^0_{a\alpha} D + Y^0_{a\alpha} H) l_a e_\alpha\,,
\eeq
where the different fields can have different 
multiplicities\footnote{This covers also the models with both down-quark
and lepton couplings, with $a=1\ldots 3$ running over quarks and $a=4\ldots6$
over leptons etc.}.
As before we define
$\phi_1\equiv D$, $\phi_2\equiv H$, $y_{1a\alpha}^0\equiv y^0_{a\alpha}$ 
and $y_{2a\alpha}^0\equiv Y^0_{a\alpha}$.
The various wave-function renormalizations are now all matrices,
and the expressions for the soft masses at 2-loops generalize to
\beq
m^2_l(\mu) = -\frac14 \, \left[
\frac{\partial^2 Z(\mu)}{\partial\ln M^2}  -
\left(\frac{\partial Z(\mu)}{\partial\ln M}\right)^2
\right]
\left|\frac{F}{M}\right|^2\,,
\eeq
and similarly for $e$.
Using the RGE for the matrix $Z_l$ (in analogy with eqn.~(\ref{matrixrun}))
this can be written as (at $\mu=M$),
\beq\label{lmassfull}
m^2_l(M) = -\frac14 \, \left[
\frac{\partial }{\partial\ln M}\Delta\gamma_l(M)-
\frac{\partial }{\partial\ln M}\gamma^<_l(\mu)\Big\vert_{\mu=M}
\right]
\left|\frac{F}{M}\right|^2\,,
\eeq
and similarly for $e$.
For completeness we display again the expression for the Higgs mass,
\beq
m^2_H(M) = \frac14 \, 
\frac{\partial }{\partial\ln M}\gamma^<_H(\mu)\Big\vert_{\mu=M}
\left|\frac{F}{M}\right|^2\,.
\eeq
Note that the second term of~(\ref{lmassfull}) is common to all the SM
fields including $H$ and $\gamma^<$ is given by the square of the low-energy
coupling $Y$. On the other hand the first term of~(\ref{lmassfull})
does not appear in the $H$ mass (since its anomalous dimension is continuous
across the threshold), and $\Delta\gamma(M)$ is given by the square of
$\tilde y(M)$.

It is now easy to evaluate these expressions. Let us do this explicitly for
the $l$ mass.
The first term of~(\ref{lmassfull}) is then
\beqa
\frac{\partial}{\partial\ln M} \Delta\gamma_l(M)_{ba}&=&
\left(-\frac2{16\pi^2}\right)\, \left[\tilde y^*_{b\alpha} 
\frac{\partial}{\partial\ln M}\tilde y_{a\alpha} + {\rm cc}
\right]\\
&=&
-\frac12\, \left(-\frac2{16\pi^2}\right)
\left[ y^*_{b\alpha}\left( 
\gamma^>_{11} y_{a\alpha} + \gamma^>_{ba}  y_{b\alpha}
+ \gamma^>_{\beta\alpha}  y_{a\beta}
+ 2\gamma^>_{12}  Y_{a\alpha}
\right)+ {\rm cc}
\right]\nonumber
\eeqa
where in the second line we omitted the tildes because the expression is already
of two-loop order.
The second term of~(\ref{lmassfull}) is,
\beqa
\frac{\partial }{\partial\ln M}\gamma^<_l(\mu)\Big\vert_{\mu=M} &=&
\left(-\frac2{16\pi^2}\right)\, \left[\tilde Y^*_{b\alpha} 
\frac{\partial}{\partial\ln M}\tilde Y_{a\alpha} + {\rm cc}
\right]\\
&=&-\frac12\, \left(-\frac2{16\pi^2}\right)
\left[ y^*_{b\alpha}\left( 
\Delta\gamma_{ba}  y_{b\alpha}
+ \Delta\gamma_{\beta\alpha}  y_{a\beta}
\right)+ {\rm cc}
\right]\,.\nonumber
\eeqa
Finally we need to substitute,
\beq
\Delta\gamma_{ba} = \left(-\frac2{16\pi^2}\right)\,
\left(y y^\dagger\right)\,
\eeq
and the analogous expression for $e$.
Here again we used the fact that $\Delta\gamma\sim \tilde y(M)^2$,
but to leading order $\tilde y=y$.

\subsection{Soft terms in the three generation model}\label{final}
We are now ready to present the soft
terms resulting from general coupling 
matrices $y_U$, $y_D$ and $y_L$. 

Note that our couplings $Y_U$ are actually the complex conjugates
of the commonly used standard-model Yukawas, which we denote by
$y_u$.
To conform with the standard notation we will therefore express the soft terms
in terms of $Y_u$ and $y_u$ (and similarly for the down and lepton couplings
with
\beqa
Y_u = Y^*_U\,,~~~ Y_d = Y^*_D\,,~~~Y_l = Y^*_L\,, \nn\\
y_u = y^*_U\,,~~~ y_d = y^*_D\,,~~~y_l = y^*_L\,. 
\eeqa

The 2-loop soft squared masses at $\mu=M$ are
\beqa
\tilde m^2_q &=
\frac{1}{(4\pi)^4}\left|\frac{F}{M}\right|^2  
\bigg\{
&
\left(3Tr\left(y_u^{\dagger}y_u \right)
-\frac{16}{3}g_3^2 - 3g_2^2 -  \frac{13}{15}g_1^2\right) y_u y_u^{\dagger}
\nn \\
&&
+\left(Tr\left(3y_d^{\dagger}y_d + y_l^{\dagger}y_l\right)
-g_3^2\frac{16}{3} - 3g_2^2 - \frac{7}{15} g_1^2\right) y_d y_d^{\dagger}
\nn \\
&&
\vphantom{\left(\frac{16}{3}\right)}
+3y_u y_u^{\dagger}y_u y_u^{\dagger} 
+ 3y_d y_d^{\dagger}y_d y_d^{\dagger}
+ y_u y_u^{\dagger} y_d y_d^{\dagger}
+ y_d y_d^{\dagger} y_u y_u^{\dagger} 
\nn\\
&&
\vphantom{\left(\frac{16}{3}\right)}+2y_u Y_u^{\dagger} Y_u y_u^{\dagger} 
+2y_d Y_d^{\dagger} Y_d y_d^{\dagger} 
-2Y_u y_u^{\dagger} y_u Y_u^{\dagger}
-2Y_d y_d^{\dagger} y_d Y_d^{\dagger}
\nn \\
&&
\vphantom{\left(\frac{16}{3}\right)}
+y_u Y_u^{\dagger} Tr\left(3y_u^{\dagger} Y_u \right)
+Y_u y_u^{\dagger} Tr\left(3Y_u^{\dagger}y_u \right)
\nn\\
&&
+y_d Y_d^{\dagger}Tr\left(3y_d^{\dagger}Y_d + y_l^{\dagger}Y_l\right)
+Y_d y_d^{\dagger}Tr\left(3Y_d^\dagger y_d + Y_l^\dagger y_l\right)
\nn\\
&&
+2N_5\left(\frac{4}{3}g_3^4 + \frac{3}{4}g_2^4 + \frac{1}{60}g_1^4\right)
1_{3\times 3}
\bigg\}
\nn \\
\eeqa

\beqa
\tilde m^2_{u_R} 
&=
\frac{1}{(4\pi)^4}\left|\frac{F}{M}\right|^2  
\bigg\{
&2\left(3Tr\left(y_u^{\dagger}y_u \right)
-\frac{16}{3}g_3^2 - 3g_2^2 - \frac{13}{15}g_1^2\right) y_u^{\dagger}y_u
\nn \\
&&
\vphantom{\left(\frac{16}{3}\right)}+6 y_u^{\dagger}y_u y_u^{\dagger}y_u 
+2y_u^{\dagger}Y_u Y_u^\dagger y_u 
+2y_u^{\dagger}Y_d Y_d^\dagger y_u 
+2y_u^{\dagger}y_d y_d^{\dagger}y_u
\nn \\
&&
-2Y_u^\dagger y_uy_u^{\dagger}Y_u 
-2Y_u^\dagger y_dy_d^{\dagger}Y_u
+2y_u^{\dagger}Y_u Tr\left(3Y_u^\dagger y_u \right)
\nn \\
&&
+2Y_u^\dagger y_u Tr\left(3y_u^{\dagger}Y_u\right)
+2\left(\frac{4}{3}g_3^4 + \frac{4}{15}\right)N_5 g_1^4 1_{3\times 3}
\bigg\}
\eeqa

\beqa
\tilde m^2_{d_R}
&=
\frac{1}{(4\pi)^4}\left|\frac{F}{M}\right|^2  
\bigg\{
&2\left(Tr\left(3y_d^{\dagger}y_d + y_l^{\dagger}y_l\right)
-\frac{16}{3}g_3^2 - 3g_2^2 -\frac{7}{15} g_1^2\right) y_d^{\dagger}y_d
\nn \\
&&
\vphantom{\left(\frac{16}{3}\right)}
+6y_d^{\dagger}y_dy_d^{\dagger}y_d
+2y_d^{\dagger}Y_u Y_u^\dagger y_d 
+2y_d^{\dagger} y_u y_u^{\dagger}y_d 
+ 2y_d^{\dagger}Y_d Y_d^\dagger y_d 
\nn \\
&&
\vphantom{\left(\frac{16}{3}\right)}
-2Y_d^\dagger y_u y_u^{\dagger}Y_d 
-2Y_d^\dagger y_d y_d^{\dagger}Y_d
\nn\\
&&
+2y_d^{\dagger}Y_d Tr\left(3Y_d^\dagger y_d + Y_l^\dagger y_l \right)
+2Y_d^\dagger y_d Tr\left(3y_d^{\dagger}Y_d + y_l^{\dagger}Y_l \right)
\nn\\
&&
+2N_5\left(\frac{4}{3}g_3^4  + \frac{1}{15}g_1^4\right) 1_{3\times 3}
\bigg\}
\nn \\
\eeqa

\beqa
\tilde m^2_L
&=
\frac{1}{(4\pi)^4}\left|\frac{F}{M}\right|^2  
\bigg\{
&\left(
Tr\left(3y_d^{\dagger}y_d + y_l^{\dagger}y_l \right) -3g_2^2 -\frac{9}{5} g_1^2
\right)y_l y_l^{\dagger}  
\nn \\
&&
\vphantom{\left(\frac{16}{3}\right)}
 +3y_l y_l^{\dagger} y_l y_l^{\dagger}   
+2y_l Y_l^{\dagger} Y_l y_l^{\dagger}     
- 2Y_l y_l^{\dagger} y_l Y_l^{\dagger}   
\nn\\\
&&
+y_l Y_l^{\dagger} Tr\left(3y_d^{\dagger}Y_d + y_l^{\dagger}Y_l\right)
+Y_l y_l^{\dagger} Tr\left(3Y_d^\dagger y_d + Y_l^\dagger y_l\right)
\nn\\\
&&
+2N_5\left(\frac{3}{4}g_2^4 + \frac{3}{20}g_1^4\right)1_{3\times 3}
\bigg\}
\eeqa

\beqa
\tilde m^2_{e_R} 
&=
\frac{1}{(4\pi)^4}\left|\frac{F}{M}\right|^2  
\bigg\{
&
2\left(
Tr\left(3y_d^{\dagger}y_d + y_l^{\dagger} y_l \right) -3g_2^2 -\frac{9}{5} g_1^2
\right)y_l^{\dagger} y_l 
\nn \\
&&
\vphantom{\left(\frac{16}{3}\right)}
+6y_l^{\dagger}y_l y_l^{\dagger}y_l
+2y_l^{\dagger}Y_l Y_l^\dagger y_l
-2Y_l^\dagger  y_l y_l^{\dagger} Y_l
\nn\\
&&
+2y_l^{\dagger}Y_l Tr\left(3Y_d^\dagger y_d + Y_l^\dagger y_l \right)
+2Y_l^\dagger y_l Tr\left(3y_d^{\dagger}Y_d + y_l^{\dagger}Y_l \right)
\nn\\
&&
+\frac{6}{5}N_5g_1^4  1_{3\times 3}
\bigg\}
\eeqa

\beqa
\tilde m^2_{H^u} 
&=
\frac{1}{(4\pi)^4}\left|\frac{F}{M}\right|^2  
\bigg\{
&-3Tr\left(
Y_u^\dagger y_u y_u^{\dagger}Y_u 
+Y_u^\dagger y_d y_d^{\dagger}Y_u
+2Y_u^\dagger Y_u y_u^{\dagger}y_u
\right)
\nn\\
&&+2N_5\left(\frac{3}{4}g_2^4 + \frac{3}{20}g_1^4\right) 
\bigg\}
\eeqa

\beqa
\tilde m^2_{H^d} 
&=
\frac{1}{(4\pi)^4}\left|\frac{F}{M}\right|^2  
\bigg\{
&
-3Tr\bigg(
Y_d^\dagger y_u y_u^{\dagger}Y_d 
+Y_d^\dagger y_d y_d^{\dagger}Y_d
+2Y_d^\dagger Y_d y_d^{\dagger}y_d
\bigg)
\nn \\
&&
-Tr\bigg(
Y_l^\dagger  y_l y_l^{\dagger} Y_l
+2Y_l^\dagger Y_l y_l^{\dagger}y_l
\bigg)
+2N_5\left(\frac{3}{4}g_2^4
 +\frac{3}{20}g_1^4\right)
\bigg\}
\nn\\
\eeqa

In addition, we give here the fully-flavored 
1-loop contributions to the soft masses,
\beqa
\delta m^2_{q_L}
&=&
-\frac{1}{(4\pi)^2}
\frac{1}{6}
\left(
y_u  y_u^{\dagger}  + y_d  y_d^{\dagger} 
\right)
 \frac{F^4}{M^6}
\\
\delta m^2_{u_R}
&=&
-\frac{1}{(4\pi)^2}
\frac{1}{3}
\left(
y_u^\dagger  y_u
\right)
 \frac{F^4}{M^6}
\\
\delta m^2_{d_R}
&=&
-\frac{1}{(4\pi)^2}
\frac{1}{3}
\left(
y_d^\dagger  y_d
\right)
 \frac{F^4}{M^6}
\\
\delta m^2_{l}
&=&
-\frac{1}{(4\pi)^2}
\frac{1}{6}
\left(
y_l y_l^{\dagger} 
\right) 
 \frac{F^4}{M^6}
\\
\delta m^2_{e^c}
&=&
-\frac{1}{(4\pi)^2}
\frac{1}{3}
\left(
y_l^\dagger  y_l
\right)
 \frac{F^4}{M^6}\,.
\eeqa

The A-terms,
i.e the coefficients of the Lagrangian terms
$L \supset (A_u)_{i,j} \tilde q_{L i} \tilde u^*_{R j} H_U 
+  (A_d)_{i,j} \tilde q_{L i} \tilde d^*_{R j} H_d
+ (A_l)_{i,j} \tilde L_{L i} \tilde e^*_{R j} H_d
$
at the scale $M$ are,
\beqa
A^*_u
&=&
-\frac{1}{16\pi^2}\left[
\left(y_u y_u^{\dagger}  + y_d y_d^{\dagger} \right)Y_u
+2Y_u\left( y_u^{\dagger}y_u \right) 
\right]\frac{F}{M}
\\
A^*_d &=& 
-\frac{1}{16\pi^2}\left[
\left(y_u y_u^{\dagger} +   y_d y_d^{\dagger}\right)Y_d
+2Y_d\left(  y_d^{\dagger} y_d \right)
\right]\frac{F}{M}
\\
A^*_l &=& 
-\frac{1}{16\pi^2}\left[
\left( y_l  y_l^{\dagger} \right)Y_l
+2Y_l\left( y_l^{\dagger} y_l\right)
\right]\,\frac{F}{M}
\eeqa

\newpage

\subsection{Explicit 2-loop Calculation}\label{explicitcal}
A a cross-check of the calculation described above, we have also
calculated the mixed $y^2 Y^2$ terms explicitly.
Since we are only interested in verifying the two loop contributions,
which are only known to leading order in $F/M^2$, we work in the limit
$F\ll M^2$, treating $F$ as an insertion. 

The scalar interaction Lagrangian is
\beqa
L_{scalar} &\supset&
- FD\bar D - F^*D^*\bar D^* - |F_D|^2 - |F_{\bar D}|^2 - |F_H|^2  - |F_l|^2 - |F_e|^2
\nn\\
&=&
-FD\bar D - F^*D^*\bar D^*- |M\bar D + yle|^2 - |MD|^2 - |Yle|^2\nn\\&&  - |YHe + yDe|^2 - |YHl + yDl|^2
\nn\\
&=&
- FD\bar D - F^*D^*\bar D^*- |M|^2 D^* D - |M|^2\bar D^*\bar D
\nn\\
&&
 -My^*\bar D l^* e^* - M^*y\bar D^* l e - (|Y|^2+|y|^2) l^*l e^* e 
\nn\\
&&
-( |Y|^2 H^*H + |y|^2 D^* D + Yy^* H D^* + Y^*y H^* D )e^*e
\nn\\
&&
-( |Y|^2 H^*H + |y|^2 D^* D + Yy^* H D^* + Y^*y H^* D )l^*l
\eeqa
and the fermion Lagrangian is
\beq
-L_{fermion} =
M\psi_D\psi_{\bar D} 
+Y (H \psi_l\psi_e + e \psi_l\psi_H + l \psi_H\psi_e)
+y (D \psi_l\psi_e + e \psi_l\psi_D + l \psi_D\psi_e)
+c.c
\eeq

For the sake of brevity we will define
\beq
\int d\varphi = \int \frac{d^4 p}{(2\pi)^4} \int \frac{d^4 k}{(2\pi)^4}
\eeq

While we are working in the $F/M^2\ll 1$ limit, it is important to 
remember that the sfermions obtain small soft mass (of order $F^4/M^2$) 
already at one loop. 
Indeed, loops of massless scalars in the calculation presented below 
lead to spurious IR divergences which manifest themselves in the fact 
that the results of the calculation appear to depend on the order of 
integration. The presence of non-vanishing scalar masses cuts off these 
divergences leading to a finite result independent of the order of 
integration. For the sake of brevity, below we choose the order of integration 
which gives the correct result even when light scalars are treated as massless.


\subsection{$H$ field soft mass squared}
The four 2-loop diagrams with two insertions of the SUSY breaking spurion and their contributions are given by,
\beqa
\includegraphics[width=0.3\textwidth]{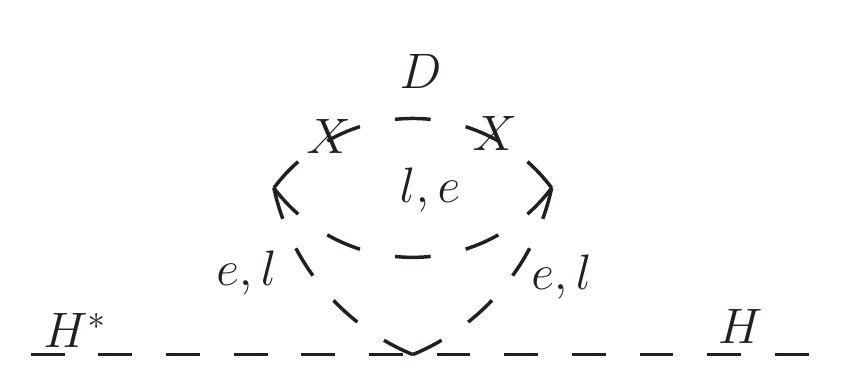} 
&=&
i|F|^2|yY|^2 \int d\varphi
\frac{M^2}{p^4(k^2-M^2)^3(p-k)^2}
\\
\includegraphics[width=0.3\textwidth]{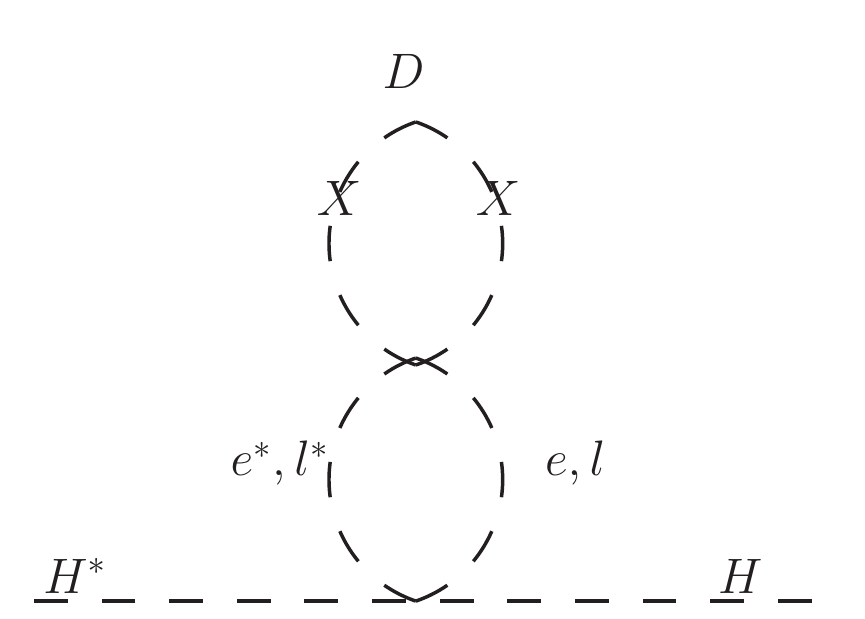} 
&=&
i|F|^2|yY|^2 \int d\varphi
\frac{1}{p^4(k^2-M^2)^3}
\\
\includegraphics[width=0.3\textwidth]{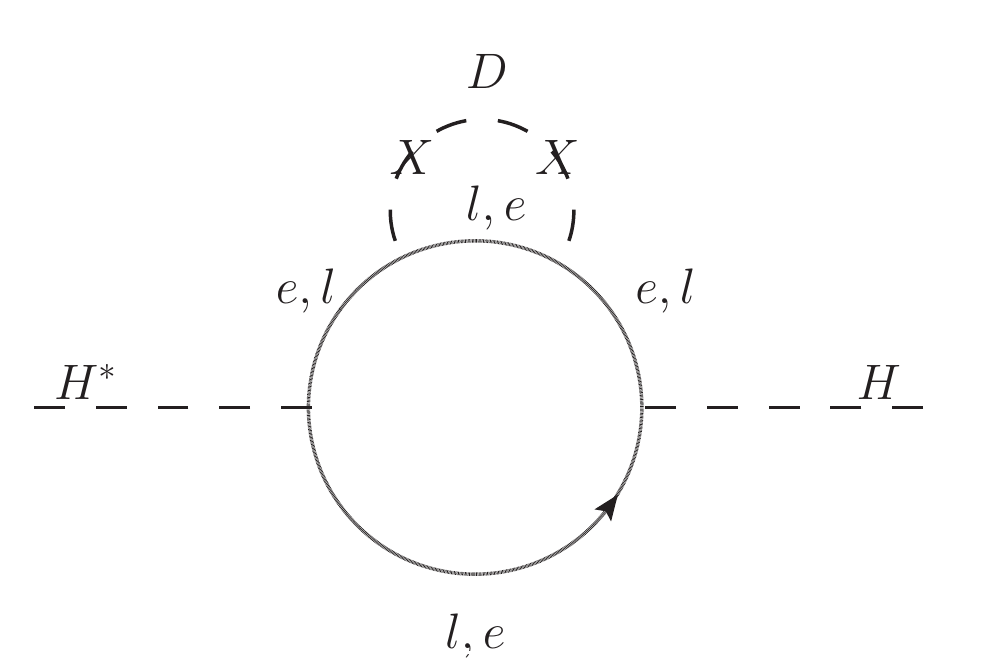} 
&=&
-i|F|^2|yY|^2 \int d\varphi
\frac{2p\cdot(p-k)}{p^4(p-k)^2 (k^2-M^2)^3} 
\\
\includegraphics[width=0.3\textwidth]{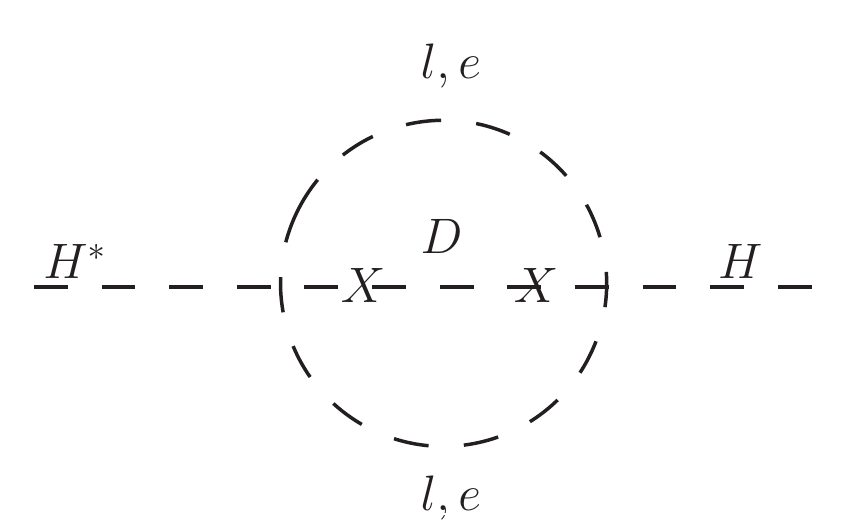} 
&=&
i|F|^2|yY|^2 \int d\varphi
\frac{1}{(k^2-M^2)^3 (p-k)^2 p^2}
\eeqa
The integrals on the right-hand side represent contributions of either $l$ or $e$ propagating in the loop. Multiplying the results by a factor of two to account for the number of fields in the loop and summing the diagrams we obtain
\beqa
I &=& 2i |Yy|^2 |F|^2
\int d\varphi
\frac{1}{p^2(k^2-M^2)^3}
\left(
\frac{M^2}{p^2(p-k)^2} 
+\frac{1}{p^2} 
+ \frac{1}{(p-k)^2}
- \frac{2p\cdot(p-k)}{p^2(p-k)^2} 
\right)
\nn\\
&=&
2i |Yy|^2 |F|^2
\int d\varphi
\frac{k^2+M^2}{p^4(k^2-M^2)^3(p-k)^2}
\label{eq:Hmass_int}
\eeqa
As discussed above, to avoid spurious IR divergences
we will choose to perform the $k$ integral, associated with 
the massive messenger loop momentum, first followed by the $p$ integral.

The resulting Higgs mass squared is 
\beqa
m^2_H = -\frac{2|Yy|^2}{(4\pi)^4}\left|\frac{F}{M}\right|^2
\eeqa
consistent with the results obtained using analytic continuation in 
\Eqref{eq:hsoft_simplified}.
\subsection{$l$ field mass squared}
The diagrams contributing to the $l$-field soft mass squared 
which contain a $|yY|^2$ term are
\beqa
\includegraphics[width=0.3\textwidth]{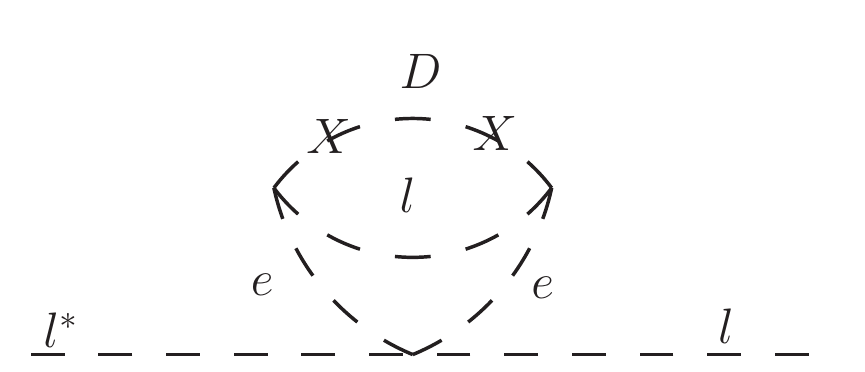} 
&=&
i|F|^2|y|^2(|y|^2+|Y|^2) \int d\varphi
\frac{|M|^2}{(k^2-M^2)^3 (p-k)^2 p^4}
\label{eq:lep1}
\\
\includegraphics[width=0.3\textwidth]{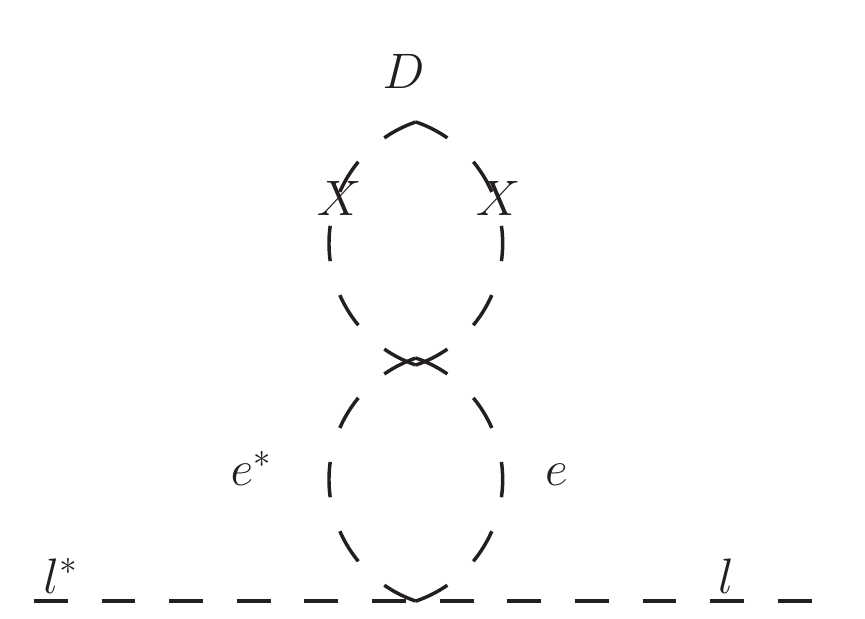} 
 &=&
i|F|^2|y|^2(|y|^2+|Y|^2) \int d\varphi
\frac{1}{(k^2-M^2)^3 p^4}
\label{eq:lep5}
\eeqa

\beqa
\includegraphics[width=0.3\textwidth]{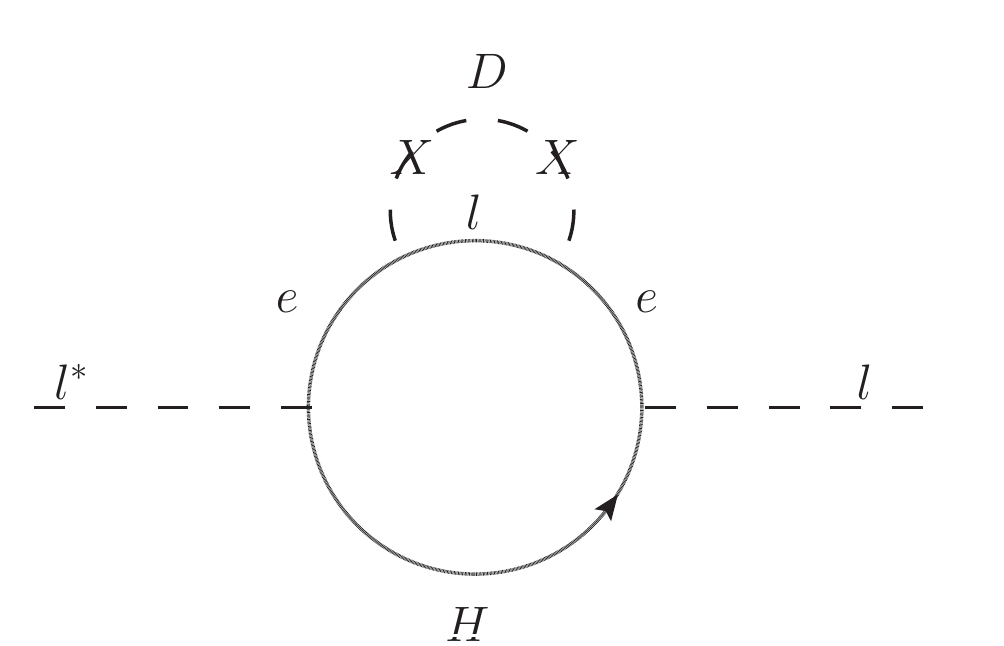} 
 &=&
-2i|F|^2|yY|^2 \int d\varphi
\frac{p\cdot(p-k)}{(k^2-M^2)^3 (p-k)^2 p^4}
\label{eq:lep16}
\\
\includegraphics[width=0.3\textwidth]{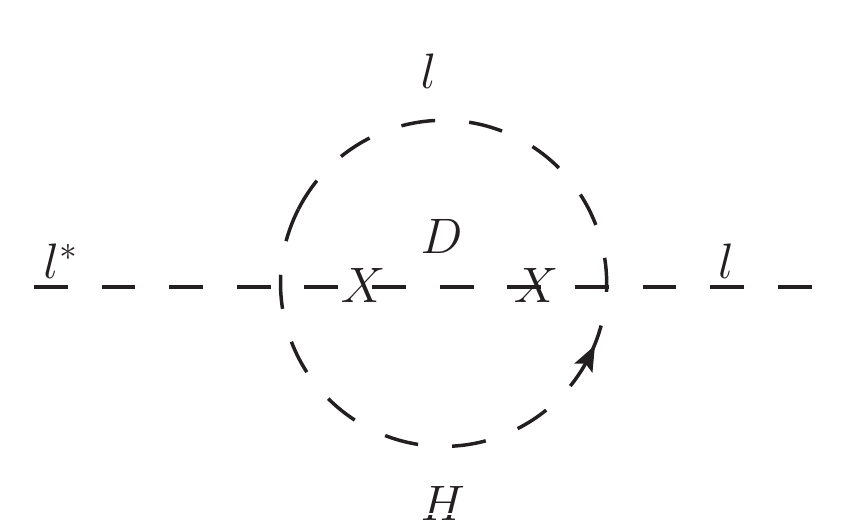} 
 &=&
2i|F|^2|yY|^2 \int d\varphi
\frac{1}{(k^2-M^2)^3 (p-k)^2 p^2}
\label{eq:lep6}
\eeqa

\beqa
\includegraphics[width=0.3\textwidth]{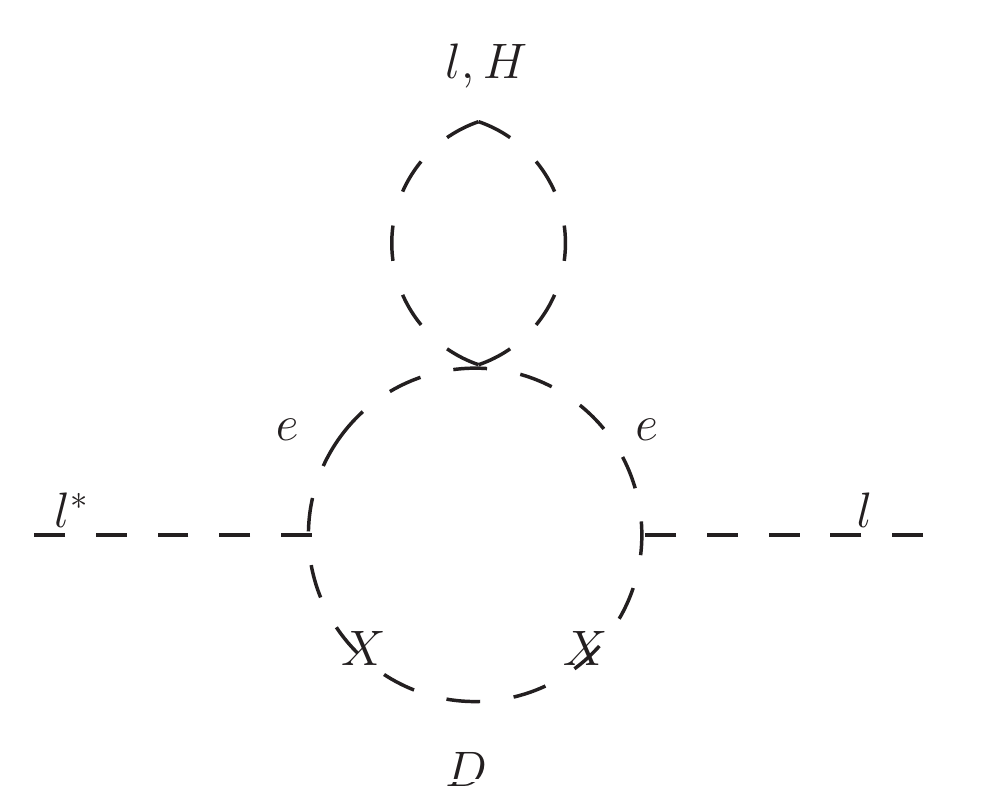} 
 &=&
i|F|^2|yM|^2(|y|^2+2|Y|^2) \int d\varphi
\frac{1}{(k^2-M^2)^3 k^4 p^2}
\label{eq:lep11}
\\
\includegraphics[width=0.3\textwidth]{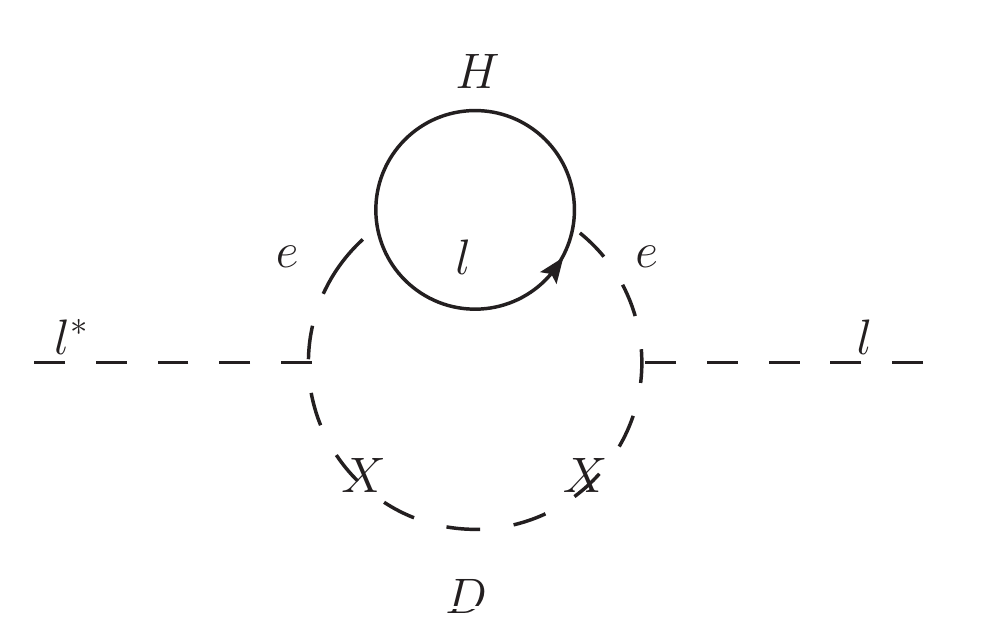} 
 &=&
-2i|F|^2|M|^2|yY|^2 \int d\varphi
\frac{p\cdot(p-k)}{(k^2-M^2)^3 k^4(p-k)^2 p^2}
\label{eq:lep14}
\eeqa
\begin{align}
\includegraphics[width=0.3\textwidth]{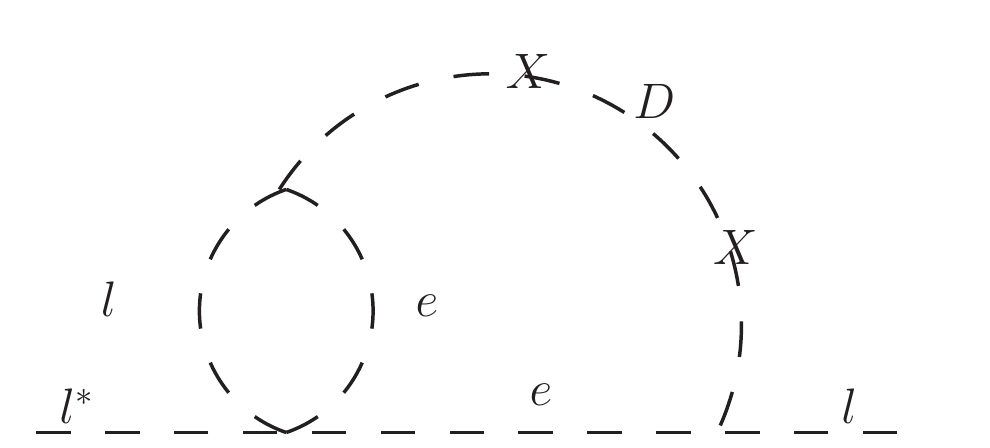} 
&+&&
\nn\\
\includegraphics[width=0.3\textwidth]{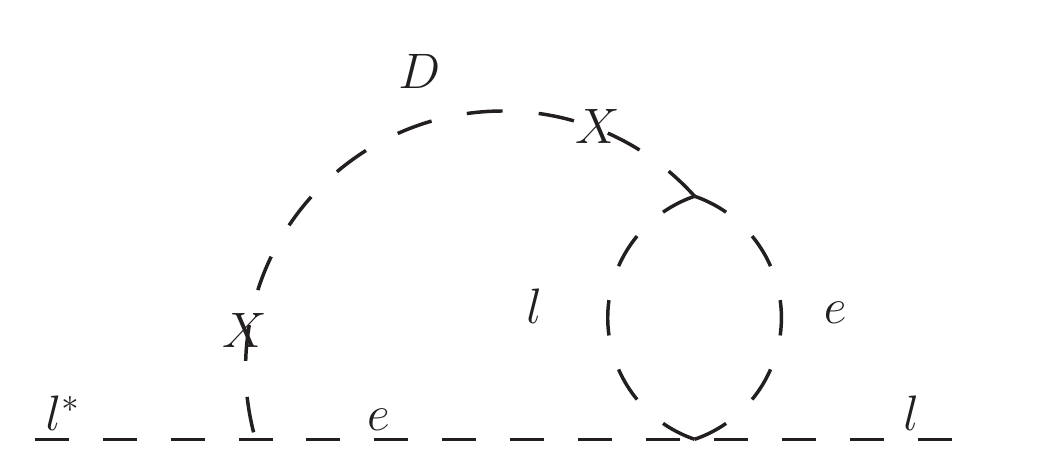} 
 &=&&2i|F|^2|y|^2(|y|^2+|Y|^2) \int d\varphi
\frac{|M|^2}{(k^2-M^2)^3 k^2 (p-k)^2 p^2}
\label{eq:lep17}
\\
\includegraphics[width=0.3\textwidth]{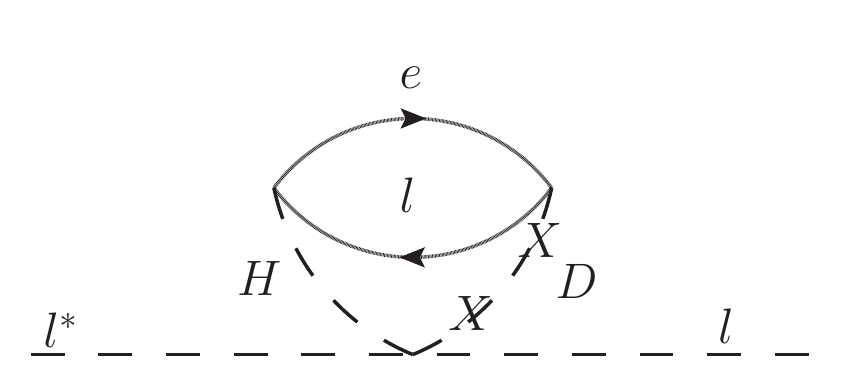} 
 &=&&-4i|F|^2|yY|^2 \int d\varphi
\frac{p\cdot(p-k)}{(k^2-M^2)^3 k^2 (p-k)^2 p^2}
\label{eq:lep19}
\\
\includegraphics[width=0.3\textwidth]{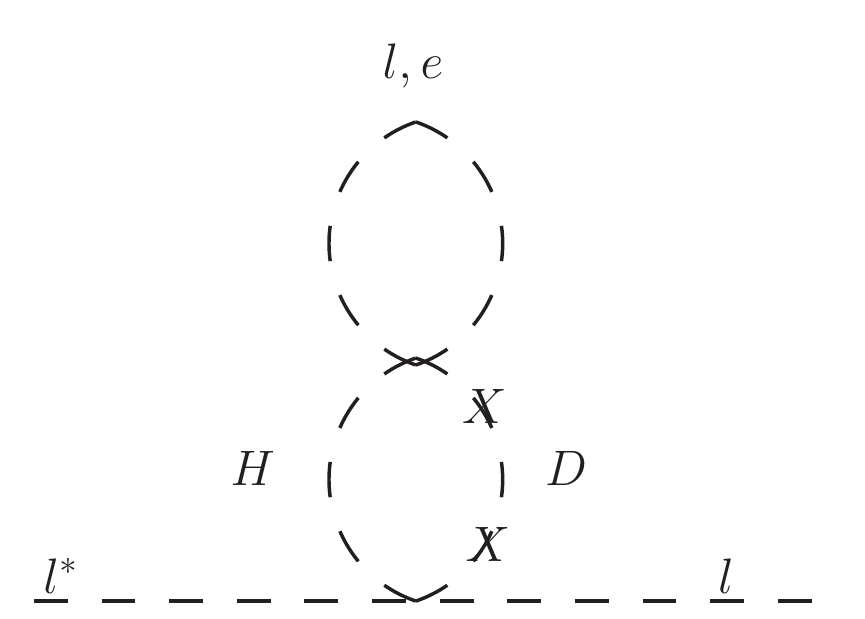} 
 &=&&
4i|F|^2|yY|^2 \int d\varphi
\frac{1}{(k^2-M^2)^3 k^2 p^2}
\label{eq:lep20}
\end{align}
We note that contributions of Feynman diagrams~\ref{eq:lep5}-\ref{eq:lep6} are identical to the diagrams contributing to the Higgs mass. The last of these diagrams, \ref{eq:lep6}, has an additional factor of two due to due to two possible choices of ``chirality'' for $D$ and $H$ propagators.
The $|yY|^2$ contribution to the $l$ soft squared mass in the 
diagrams of 
Eqs.~\ref{eq:lep5}-\ref{eq:lep20}
can be written as a sum of three integrals,
\begin{equation}
\begin{split}
{\cal I}&=\int d\varphi
 \frac{k^2+M^2}{(k^2-M^2)^3 (p-k)^2 p^4}=\frac{1}{(4\pi)^4M^2}\\
{\cal II}& =  3 \int d\varphi
\frac{k^2+M^2}{k^2(k^2-M^2)^3 (p-k)^2 p^2}=-\frac{3}{(4\pi)^4 M^2}
\\
{\cal III}&=\int d\varphi 
\frac{(p-k)^2 - p^2}{(k^2-M^2)^3 (p-k)^2 p^2}
 \left( \frac{2}{k^2} + \frac{M^2}{k^4}\right) =0\,.
\end{split}
\end{equation}

Summing all contributions, the $|yY|^2$ part of the $l$ 
soft squared mass reads
\beq
m^2_l\bigg|_{|yY|^2} = \frac{2|Yy|^2}{(4\pi)^4}\left|\frac{F}{M}\right|^2
\eeq
consistent with the results obtained in the revised analytic continuation
in \Eqref{eq:lsoft_simplified}.




\end{document}